\documentclass[aps,prl,preprint, superscriptaddress]{revtex4-1}

\usepackage{graphicx}  
\usepackage{bm} 
\usepackage{amssymb}  
\usepackage{amsmath}
\usepackage{mhchem}
\usepackage{mathptmx}
\usepackage{siunitx}
\usepackage{afterpage}
\usepackage{color}

\begin{document}
\widetext


\title{Polarized Raman spectroscopy study of metallic \ce{(Sr_{1-x}La_{x})_{3}Ir2O7}: a consistent picture of disorder-interrupted unidirectional charge order}


\author{Wencan Jin}
\affiliation{Department of Physics, University of Michigan, Ann Arbor, Michigan 48109, USA}

\author{Siwen Li}
\affiliation{Department of Physics, University of Michigan, Ann Arbor, Michigan 48109, USA}

\author{Jianpeng Liu}
\affiliation{Kavli Institute of Theoretical Physics, University of California, Santa Barbara, California 93106, USA}

\author{Qiang Han}
\affiliation{Department of Physics and Astronomy, Rutgers University, Piscataway, New Jersey 08854, USA}

\author{Zach Porter}
\affiliation{Materials Department, University of California, Santa Barbara, California 93106, USA}

\author{Christi Peterson}
\affiliation{Materials Department, University of California, Santa Barbara, California 93106, USA}

\author{Julian Schmehr}
\affiliation{Materials Department, University of California, Santa Barbara, California 93106, USA}

\author{Ibrahim Boulares}
\affiliation{Department of Physics, University of Michigan, Ann Arbor, Michigan 48109, USA}

\author{Kai Sun}
\affiliation{Department of Physics, University of Michigan, Ann Arbor, Michigan 48109, USA}

\author{Roberto Merlin}
\affiliation{Department of Physics, University of Michigan, Ann Arbor, Michigan 48109, USA}

\author{Stephen D. Wilson}
\affiliation{Materials Department, University of California, Santa Barbara, California 93106, USA}

\author{Liuyan Zhao} 
\email{lyzhao@umich.edu}
\affiliation{Department of Physics, University of Michigan, Ann Arbor, Michigan 48109, USA} 


\begin{abstract}
We have used rotational anisotropic polarized Raman spectroscopy to study the symmetries, the temperature and the doping dependence of the charge ordered state in metallic \ce{(Sr_{1-x}La_{x})_{3}Ir2O7}. Despite the fact that the Raman probe size is greater than the charge ordering length, we establish that the charge ordering breaks the fourfold rotational symmetry of the underlying tetragonal crystal lattice into twofold, as well as the translational symmetry, and forms short-range domains with \SI{90}{\degree} rotated charge order wave vectors, as soon as the charge order sets in below $T_\mathrm{CO} = \sim$ \SI{200}{K} and across the doping-induced insulator metal transition. We observe that this charge order mode frequency remains nearly constant over a wide temperature range and up to the highest doping level. These features above are highly reminiscent of the ubiquitous unidirectional charge order in underdoped high-$T_\mathrm{C}$ copper-oxide-based superconductors (cuprates). We further resolve that the charge order damping rate diverges when approaching $T_\mathrm{CO}$ from below and increases significantly as increasing the La-doping level, which resembles the behaviors for a disorder-interrupted ordered phase and has not been observed for the charge order in cuprates.
\end{abstract}

\maketitle

Chemical doping, a widely used experimental technique in materials science and condensed matter physics research, opens up a fruitful pathway towards discoveries of new materials and exotic quantum phases. However, doping inevitably introduces quenched disorder that often conceals the underlying physics, making it much more challenging to detect and understand their ground states. For example, it has taken a decade to resolve the symmetries of charge order in underdoped cuprates \cite{tranquada1995evidence, chang2012direct, ghiringhelli2012long, comin2015symmetry, comin2015broken, comin2016resonant}, and remains elusive to disentangle its relationship to other cuprate phenomena \cite{fradkin2012high, fradkin2015colloquium} including the Mott insulating Heisenberg antiferromagnetic (AFM) state, the pseudogap behavior, the $q = 0$ magnetic multipolar order, and the high-$T_\mathrm{C}$ superconductivity.

Theoretically, it has been known for a long time that an infinitesimal amount of randomness is sufficient to suppress the long-range order with broken continuous symmetries in three dimensions (3D) or broken continuous or discrete symmetries in two dimensions (2D), turning the ordered phase into a set of short-range ordered domains or clusters \cite{imry1975random}. More recently, such effects have been predicted to lead to intriguing phenomena, such as vestigial order \cite{nie2014quenched}. In particular, a nematic vestigial order in hole-doped cuprates -- a $q = 0$ point-group symmetry-breaking order -- has been proposed as the result of partial melting of the unidirectional charge order.

Experimentally, probing the intrinsic properties, e.g., symmetries, of disorder-interrupted ordered states has been very challenging as it requires one to overcome the difficulty of cluster-average effects. So far, there have been two types of advanced experimental tools suitable for addressing this challenge. The first type includes local probes such as scanning tunneling spectroscopy (STS) mapping and transmission electron microscopy (TEM), which are sensitive to the local electronic states and have probe sizes smaller than the short-range ordering length. STS mapping requires a careful analysis to distinguish electronic order from the disorder-induced quasiparticle interference pattern \cite{da2014ubiquitous, arguello2015quasiparticle}, whereas TEM probes the tiny lattice distortions that emerge from the electronic order \cite{el2018nature}. The second type includes resonant x-ray diffraction \cite{comin2015broken, comin2016resonant} whose diffraction peaks can capture the structure factors for domain mesostructures of short-range orders, and nuclear magnetic resonance spectroscopy  \cite{wu2011magnetic, wu2015incipient} whose line shapes contain the information of local symmetries of short-range orders, despite the fact that their probe scales are larger than the domain size.

In this work, we present a study of charge order in lightly La-doped \ce{Sr3Ir2O7}, using rotational anisotropic (RA) polarized Raman spectroscopy. We show that even though the Raman probe size exceeds the cluster size, the symmetry point group of the charge order can be uniquely determined as $D_{2h}$, a subgroup of the lattice point group $D_{4h}$, based on Raman selection rules. We also identify that this twofold rotational ($C_2$ symmetry about \textit{c} axis) charge order exhibits a charge density modulation along the Ir-O-Ir bond, the diagonal direction of the expanded unit cell of \ce{Sr3Ir2O7}. Moreover, we show from the RA pattern analysis that two types of short-range charge order domains develop in the vicinity of La-dopants, whose charge order orientations are \SI{90}{\degree} rotated from each other. Finally, we draw from temperature and doping dependence data that the frequency of the mode associated with the charge order remains nearly constant over a wide temperature range and up to the highest available doping levels, while its damping rate diverges as one approaches the transition temperature $T_\mathrm{CO}$ from below, and it increases significantly at higher doping levels. The frequency behavior is reminiscent of the unidirectional charge order observed in underdoped cuprates, whereas the divergence towards $T_\mathrm{CO}$ and the increase upon doping of the damping rate resemble those of other disorder-interrupted ordered states, which have not been observed in cuprates. Our experiment also show that Raman scattering possesses the ability to retrieve single cluster information using a domain-averaged signal acquired with large probe sizes.

Unlike the single-layer perovskite iridate \ce{Sr2IrO4}, believed to be the spin-orbit coupled (SOC) cuprate analogue, which hosts the $J_{eff} = 1/2$ Mott insulating Heisenberg AFM state \cite{kim2008novel, kim2009phase}, pseudogap behavior \cite{kim2014fermi, de2015collapse, cao2016hallmarks, battisti2017universality}, magnetic quadrupolar order \cite{zhao2016evidence, jeong2017time} and a \textit{d}-wave gap \cite{yan2015electron, kim2016observation}, the bilayer variant, \ce{(Sr_{1-x}La_{x})_{3}Ir2O7}, is simpler in that it only exhibits an electron-doping-induced insulator metal transition (IMT) near $x_c \sim 0.03$ \cite{li2013tuning, hogan2015first} and a charge density wave (CDW)-like instability across IMT \cite{chu2017charge}. This makes the \ce{(Sr_{1-x}La_{x})_{3}Ir2O7} system a neat platform to study the disorder-interrupted charge ordering in a correlated metal without the complication of other competing phases as cuprates do and near IMT that cuprates cannot access. Previous work has confirmed the phase purity of our \ce{(Sr_{1-x}La_{x})_{3}Ir2O7} crystals with \textit{x} = 0 - 0.075, and revealed subtle structural distortions of the \ce{IrO6} octahedra pair tilting off the \textit{c} axis that lower the crystal point group from tetragonal $D_{4h}$ to monoclinic $C_{2h}$ with one remaining mirror plane normal to the in-plane crystal axis \textit{b} and a $C_2$ axis about \textit{b} axis \cite{hogan2016structural, hogan2017doping}. Moreover, despite suggesting the possibility of a charge order in metallic \ce{(Sr_{1-x}La_{x})_{3}Ir2O7} in Ref. \cite{chu2017charge}, much remains unknown about this charge order, such as its full symmetries, domain structures, doping, and temperature dependence, which are, however, necessary for understanding the nature of this charge order and comparing it with that in cuprates.

Raman measurements were performed in the backscattering geometry, using the 514 nm line of an argon-ion laser and a Dilor XY triple monochromator spectrometer. The incident optical fluence was maintained at 1.2 mW over a beam diameter of 20 $\mu$m that is smaller than the size of reported monoclinic structural domains \cite{hogan2017doping}. To perform the RA polarized Raman measurements (Fig. 1(a)), the incident ($E_I$, solid arrow) and scattered ($E_S$, dashed arrow) electric fields were kept either parallel or perpendicular to each other, and rotated by an angle $\phi$ with respect to the crystal axis \textit{a}. To avoid any background anisotropy from the setup, we used strain-free optical components to minimize light depolarization, and rotated the selected scattered polarization back to vertical to ensure a constant collection efficiency.

Figure 1(b) shows Raman spectra in the four high-symmetry channels for the tetragonal lattice ($D_{4h}$) of \ce{(Sr_{1-x}La_{x})_{3}Ir2O7} with \textit{x} = 0.075 at 80 K, namely, parallel and crossed electric fields along the crystal axis \textit{a} ($\phi$ = \SI{0}{\degree}) and the diagonal direction ($\phi$ = \SI{45}{\degree}) that are denoted as \textit{aa}, \textit{ab}, $a^{\prime}a^{\prime}$ and $a^{\prime}b^{\prime}$ respectively. We clearly identify all the phonon modes of frequencies greater than \SI{100} {cm^{-1}} to be of either $A_{1g}$ or $B_{2g}$ symmetry of the $D_{4h}$ point group. The mode at around \SI{330} {cm^{-1}} seemingly violates the $D_{4h}$ selection rule by appearing in both $A_{1g}$ and $B_{2g}$ configurations. However, a careful analysis shows that this band consists of two Raman modes, of $A_{1g}$ and $B_{2g}$ symmetry at, respectively, 330.1 $\pm$ \SI{0.7} and 322.8 $\pm$ \SI{0.8} {cm^{-1}}, whose frequency separation of \SI{7.3} {cm^{-1}} is well beyond our instrument resolution of \SI{1.7} {cm^{-1}}. Here, the assigned $D_{4h}$ point group for \ce{(Sr_{1-x}La_{x})_{3}Ir2O7} single crystals, though consistent with the Raman literature \cite{gretarsson2016two}, is of higher symmetry than the recently refined monoclinic $C_{2h}$ point group, possibly because the structural distortion is so subtle that high-frequency ($>$ \SI{100} {cm^{-1}}) phonons are not appreciably affected.

In sharp contrast to the high frequency phonons, the low frequency mode at 25.9 $\pm$ \SI{0.2} {cm^{-1}}, having the same energy as the reported CDW amplitude mode \cite{chu2017charge}, shows up in both $A_{1g}$ and $B_{2g}$ channels, indicating that its ordered ground state has a lower symmetry point group than $D_{4h}$. By surveying all of the Raman tensors of the $D_{4h}$ subgroups that have finite tensor elements in the \textit{aa}, \textit{ab}, and $a^{\prime}a^{\prime}$ channels but is fully suppressed in the $a^{\prime}b^{\prime}$ one (see Supplementary Section 1 \cite{Supplementary}), we found that this mode belongs to the A ($A_g$, $A_1$, and A) symmetry of the orthorhombic subgroups ($D_{2h}$, $C_{2v}$, and $D_2$, respectively). Considering the fact that this mode is only present at low temperatures (see Supplementary Section 2 \cite{Supplementary}) while second harmonic generation shows no changes from room to low temperature \cite{hogan2017doping}, we ruled out the two noncentrosymmetric candidates ($C_{2v}$ and $D_2$) and uniquely identified the point group $D_{2h}$ for the emergent order. 

There are two important remarks regarding this point-group symmetry-breaking order. First, symmetry breaking from $D_{4h}$ to $D_{2h}$ is consistent with both the nematic order ($q = 0$) and the unidirectional charge order ($q \neq 0$). To distinguish between the two, we examined if broken translational symmetry accompanies the point-group symmetry breaking. Assuming $q = 0$ across the transition, based on the compatibility relation of the point groups, the $A_{1g}$ irreducible representation (irrep) of $D_{2h}$ traces back to either the $A_{1g}$ or $B_{2g}$ irrep of $D_{4h}$, both of which are Raman active and thus should be present at room temperature. The fact that this mode only appears at low temperatures suggests that the phase transition involves translational symmetry breaking in addition to the rotational symmetry breaking, which is evidence for unidirectional charge order, instead of nematic order. Second, the presence of the new mode in the \textit{ab} but not the $a^{\prime}b^{\prime}$ channel suggests that the unidirectional charge order modulation is along the diagonal direction, i.e., the Ir-O-Ir bond direction, which preserves the two diagonal mirrors and $C_2$ about the \textit{c} axis, but breaks the two vertical mirrors normal to the crystal axes \textit{a} and \textit{b} of the $D_{4h}$ crystal lattice. This is in stark contrast to the reported lattice distortion that breaks both diagonal mirrors and preserves one of the two vertical mirrors and a $C_2$ axis along the \textit{b} axis, excluding the possibility that this mode results from the reported monoclinic structural distortion.

For the unidirectional charge order, quenched disorder can lead to two possible domain mesostructures in the form of (i) short-range ordered domains with the same rotational but interrupted translational order, therefore sharing one charge order wave vector, and (ii) clusters with both the long-range rotational and translational ordering destroyed, leading to two types of wave vectors that are \SI{90}{\degree} rotated from each other \cite{nie2014quenched}. To examine these disorder-induced clusters, we performed RA Raman spectroscopy measurements with an angular resolution of \SI{10}{\degree}, and extracted the peak intensities of the charge order mode by fitting the Raman spectra with a one oscillator function of the form $I(\omega)=Im(\omega^2-\omega_0^2-i\omega\Gamma)\cdot \kappa^2 \cdot[\beta(\omega)+1]$, where $\omega_0$, $\Gamma$, $\kappa$, and $\beta(\omega)$ are frequency, damping rate, coupling constant, and Bose-Einstein factor, respectively. \cite{burns1970raman, travaglini1983cdw}.

Figure 2 shows the polar plots of the fitted peak intensity, $I(\omega_0)$, as a function of angle $\phi$ for the parallel (top) and the crossed (bottom) configurations for \ce{(Sr_{1-x}La_{x})_{3}Ir2O7} with \textit{x} = 0.075. The nonzero background in the RA pattern for the parallel geometry is consistent with a rotational symmetry lower than that of the $D_{4h}$ point group, similar to the discussion in Fig. 1(b). The patterns for both channels exhibit fourfold rotational symmetry, which is, however, incompatible with the unidirectional charge order domains with the same orientation. Instead, both patterns can be well fitted by a linear superposition of Raman signals from two domains with \SI{90}{\degree} rotated unidirectional charge orders, as shown in the second and third columns in Fig. 2. No observable spatial variations were detected when scanning across the sample surface with our 20-$\mu$m-diameter probe beam, suggesting that the average domain size of the unidirectional charge order is well below 20 $\mu$m, much smaller than that of the recently reported monoclinic structural domains. This drastically different charge order and monoclinic structural domain sizes, as well as the lower symmetry of the charge order mode than any detectable phonon modes, suggest a weak coupling between the charge order and the monoclinic lattice distortion.

Having established the low-temperature characteristics, we tracked the temperature dependence of the unidirectional charge order mode to further understand its order parameter. Figure 3(a) shows Raman spectra and their one-oscillator fits as a function of temperature obtained from \ce{(Sr_{1-x}La_{x})_{3}Ir2O7} with x = 0.032. We further quantify the temperature dependence by plotting the fitted mode parameters, including peak intensity $I(\omega_0)$ and coupling constant $\kappa$, damping rate $\Gamma$ and frequency $\omega_0$, against the temperature in Figs. 3(b)-3(d). $I(\omega_0)$ clearly exhibits a linear dependence on $T_\mathrm{CO}-T$ that gives the critical temperature $T_\mathrm{CO}$ = 220.1 $\pm$ 1.4 K, while $\kappa$ scales as $(T_\mathrm{CO}-T)^\varsigma$ with $\varsigma$ = 0.32 $\pm$ 0.04, a value that departs significantly from the mean-field value of 0.5 for conventional CDWs. Going beyond Ref. \cite{chu2017charge}, the large increase of $\Gamma$ when approaching $T_\mathrm{CO}$ from below shows a diverging behavior that is well described by $\Gamma = \Gamma_0 (T_\mathrm{CO}-T)^{-1/2}$, in agreement with the expectation of a divergence due to anomalous fluctuations near a phase transition in quasi-low-dimensional systems \cite{travaglini1983cdw}. Different from what was reported in Ref. \cite{chu2017charge}, the weak softening of less than $10\%$ of $\omega_0$ sharply contrasts to that of the point-group symmetry-preserving CDWs in pristine samples where the CDW amplitude mode approaches zero frequency at $T_\mathrm{CO}$ \cite{tsang1976raman, gor2012charge}. However, it is highly reminiscent to that of the rotational-symmetry-breaking CDWs in doped compounds such as cuprates \cite{torchinsky2013fluctuating, hinton2013new} and \ce{K_{0.3}MoO3} \cite{travaglini1983cdw}. This unconventional temperature-dependent behavior evokes that expected for a disorder-interrupted point-group symmetry-breaking order, and invites further theoretical investigations to distinguish disorder effects from intrinsic effects due to strong electron correlations.

By carrying out temperature-dependent Raman measurements and symmetry analysis similar to Figs. 1-3 for four La-doping levels spanning from pristine to the highest available doping level, we confirmed that both types of short-range domains with \SI{90}{\degree} rotated unidirectional charge order orientations coexist at mesoscopic scales as soon as the unidirectional charge order sets in around 200 K and across the IMT at $x_c\sim0.03$. Such a small disorder threshold for inducing the short-range domains is consistent with the quasi-2D electronic structure of \ce{(Sr_{1-x}La_{x})_{3}Ir2O7}. We further compare the charge order mode parameters against the La-doping level in Fig. 4. The frequency $\omega_0$ of the charge order mode barely changes while the damping rate $\Gamma$ increases by $50\%$, as the La doping varies from \textit{x} = 0.032 to 0.075. The constant charge order excitation energy, together with the nearly constant $T_\mathrm{CO}$, suggests that the important ingredient at the Fermi surface for developing unidirectional charge order remains nearly unchanged across doping levels in metallic \ce{(Sr_{1-x}La_{x})_{3}Ir2O7}, one possibility being the wave vector connecting the neighboring electron pockets \cite{de2014coherent} and its joint charge density of states. The significantly increased damping rate is consistent with defect-enhanced fluctuations in quasi-2D electron systems.

In conclusion, careful symmetry analysis on the RA polarized Raman spectra of metallic \ce{(Sr_{1-x}La_{x})_{3}Ir2O7} enabled us to resolve the symmetry point group and the wave vector orientation of the unidirectional charge order and establish the presence of both types of short-range domains with \SI{90}{\degree} rotated charge order wave vectors, despite the challenge of the probe beam being greater than the charge ordering length scale. The symmetry and the constant frequency of this charge order, as well as its emergence from doping a Mott insulator, are highly reminiscent of the ubiquitous charge order in underdoped curpates. Meanwhile, the doping range near IMT of \ce{(Sr_{1-x}La_{x})_{3}Ir2O7}, inaccessible in cupates, provides us with the unique opportunity to resolve the divergent temperature dependence towards $T_\mathrm{CO}$ and the strong increase upon doping for the damping rate of this charge order, which highly resembles those for a disorder-interrupted ordered phase. Furthermore, our RA polarized Raman technique of retrieving single-domain information out of a domain-averaged signal is applicable to a broader class of point-group symmetry-breaking orders with defect interruptions.\\

\noindent\textbf{ACKNOWLEDGEMENTS} \\
We acknowledge useful discussions with E. Gull, S. Coh, V. Madhavan and D. Hsieh. We also thank L. Chen and L. Zhang for technical assistance. L. Z. acknowledges support by NSF CAREER Award No. DMR-1749774 (S. L. and L. Z.). S. D. W. acknowledges support by NSF Award No. DMR-1505549 and by ARO Award No. W19NF-16-1-0361 (Z. P. and C. P.). K. S. acknowledges support by NSF Award No. NSF-EFMA-1741618 and the Alfred P. Sloan Foundation.\\

\newpage

\bibliographystyle{apsrev4-1}
\nocite{apsrev41Control}
\bibliography{Maintext.bib} 

\begin{thebibliography}{39}%
\makeatletter
\providecommand \@ifxundefined [1]{%
 \@ifx{#1\undefined}
}%
\providecommand \@ifnum [1]{%
 \ifnum #1\expandafter \@firstoftwo
 \else \expandafter \@secondoftwo
 \fi
}%
\providecommand \@ifx [1]{%
 \ifx #1\expandafter \@firstoftwo
 \else \expandafter \@secondoftwo
 \fi
}%
\providecommand \natexlab [1]{#1}%
\providecommand \enquote  [1]{``#1''}%
\providecommand \bibnamefont  [1]{#1}%
\providecommand \bibfnamefont [1]{#1}%
\providecommand \citenamefont [1]{#1}%
\providecommand \href@noop [0]{\@secondoftwo}%
\providecommand \href [0]{\begingroup \@sanitize@url \@href}%
\providecommand \@href[1]{\@@startlink{#1}\@@href}%
\providecommand \@@href[1]{\endgroup#1\@@endlink}%
\providecommand \@sanitize@url [0]{\catcode `\\12\catcode `\$12\catcode
  `\&12\catcode `\#12\catcode `\^12\catcode `\_12\catcode `\%12\relax}%
\providecommand \@@startlink[1]{}%
\providecommand \@@endlink[0]{}%
\providecommand \url  [0]{\begingroup\@sanitize@url \@url }%
\providecommand \@url [1]{\endgroup\@href {#1}{\urlprefix }}%
\providecommand \urlprefix  [0]{URL }%
\providecommand \Eprint [0]{\href }%
\providecommand \doibase [0]{http://dx.doi.org/}%
\providecommand \selectlanguage [0]{\@gobble}%
\providecommand \bibinfo  [0]{\@secondoftwo}%
\providecommand \bibfield  [0]{\@secondoftwo}%
\providecommand \translation [1]{[#1]}%
\providecommand \BibitemOpen [0]{}%
\providecommand \bibitemStop [0]{}%
\providecommand \bibitemNoStop [0]{.\EOS\space}%
\providecommand \EOS [0]{\spacefactor3000\relax}%
\providecommand \BibitemShut  [1]{\csname bibitem#1\endcsname}%
\let\auto@bib@innerbib\@empty
\bibitem [{\citenamefont {Tranquada}\ \emph {et~al.}(1995)\citenamefont
  {Tranquada}, \citenamefont {Sternlieb}, \citenamefont {Axe}, \citenamefont
  {Nakamura},\ and\ \citenamefont {Uchida}}]{tranquada1995evidence}%
  \BibitemOpen
  \bibfield  {author} {\bibinfo {author} {\bibfnamefont {J.}~\bibnamefont
  {Tranquada}}, \bibinfo {author} {\bibfnamefont {B.}~\bibnamefont
  {Sternlieb}}, \bibinfo {author} {\bibfnamefont {J.}~\bibnamefont {Axe}},
  \bibinfo {author} {\bibfnamefont {Y.}~\bibnamefont {Nakamura}}, \ and\
  \bibinfo {author} {\bibfnamefont {S.}~\bibnamefont {Uchida}},\ }\bibfield
  {title} {\enquote {\bibinfo {title} {Evidence for stripe correlations of
  spins and holes in copper oxide superconductors},}\ }\href@noop {} {\bibfield
   {journal} {\bibinfo  {journal} {Nature}\ }\textbf {\bibinfo {volume}
  {375}},\ \bibinfo {pages} {561--563} (\bibinfo {year} {1995})}\BibitemShut
  {NoStop}%
\bibitem [{\citenamefont {Chang}\ \emph {et~al.}(2012)\citenamefont {Chang},
  \citenamefont {Blackburn}, \citenamefont {Holmes}, \citenamefont
  {Christensen}, \citenamefont {Larsen}, \citenamefont {Mesot}, \citenamefont
  {Liang}, \citenamefont {Bonn}, \citenamefont {Hardy}, \citenamefont
  {Watenphul} \emph {et~al.}}]{chang2012direct}%
  \BibitemOpen
  \bibfield  {author} {\bibinfo {author} {\bibfnamefont {J.}~\bibnamefont
  {Chang}}, \bibinfo {author} {\bibfnamefont {E.}~\bibnamefont {Blackburn}},
  \bibinfo {author} {\bibfnamefont {A.}~\bibnamefont {Holmes}}, \bibinfo
  {author} {\bibfnamefont {N.}~\bibnamefont {Christensen}}, \bibinfo {author}
  {\bibfnamefont {J.}~\bibnamefont {Larsen}}, \bibinfo {author} {\bibfnamefont
  {J.}~\bibnamefont {Mesot}}, \bibinfo {author} {\bibfnamefont
  {R.}~\bibnamefont {Liang}}, \bibinfo {author} {\bibfnamefont
  {D.}~\bibnamefont {Bonn}}, \bibinfo {author} {\bibfnamefont {W.}~\bibnamefont
  {Hardy}}, \bibinfo {author} {\bibfnamefont {A.}~\bibnamefont {Watenphul}},
  \emph {et~al.},\ }\bibfield  {title} {\enquote {\bibinfo {title} {Direct
  observation of competition between superconductivity and charge density wave
  order in \ce{YBa2Cu3O_{6.67}}},}\ }\href@noop {} {\bibfield  {journal}
  {\bibinfo  {journal} {Nature Physics}\ }\textbf {\bibinfo {volume} {8}},\
  \bibinfo {pages} {871--876} (\bibinfo {year} {2012})}\BibitemShut {NoStop}%
\bibitem [{\citenamefont {Ghiringhelli}\ \emph {et~al.}(2012)\citenamefont
  {Ghiringhelli}, \citenamefont {Le~Tacon}, \citenamefont {Minola},
  \citenamefont {Blanco-Canosa}, \citenamefont {Mazzoli}, \citenamefont
  {Brookes}, \citenamefont {De~Luca}, \citenamefont {Frano}, \citenamefont
  {Hawthorn}, \citenamefont {He} \emph {et~al.}}]{ghiringhelli2012long}%
  \BibitemOpen
  \bibfield  {author} {\bibinfo {author} {\bibfnamefont {G.}~\bibnamefont
  {Ghiringhelli}}, \bibinfo {author} {\bibfnamefont {M.}~\bibnamefont
  {Le~Tacon}}, \bibinfo {author} {\bibfnamefont {M.}~\bibnamefont {Minola}},
  \bibinfo {author} {\bibfnamefont {S.}~\bibnamefont {Blanco-Canosa}}, \bibinfo
  {author} {\bibfnamefont {C.}~\bibnamefont {Mazzoli}}, \bibinfo {author}
  {\bibfnamefont {N.}~\bibnamefont {Brookes}}, \bibinfo {author} {\bibfnamefont
  {G.}~\bibnamefont {De~Luca}}, \bibinfo {author} {\bibfnamefont
  {A.}~\bibnamefont {Frano}}, \bibinfo {author} {\bibfnamefont
  {D.}~\bibnamefont {Hawthorn}}, \bibinfo {author} {\bibfnamefont
  {F.}~\bibnamefont {He}},  \emph {et~al.},\ }\bibfield  {title} {\enquote
  {\bibinfo {title} {Long-range incommensurate charge fluctuations in
  ${(Y,Nd)}{Ba}_{2}{Cu}_{3}{O}_{6+x}$},}\ }\href@noop {} {\bibfield  {journal}
  {\bibinfo  {journal} {Science}\ }\textbf {\bibinfo {volume} {337}},\ \bibinfo
  {pages} {821--825} (\bibinfo {year} {2012})}\BibitemShut {NoStop}%
\bibitem [{\citenamefont {Comin}\ \emph
  {et~al.}(2015{\natexlab{a}})\citenamefont {Comin}, \citenamefont {Sutarto},
  \citenamefont {He}, \citenamefont {da~Silva~Neto}, \citenamefont {Chauviere},
  \citenamefont {Fra{\~n}o}, \citenamefont {Liang}, \citenamefont {Hardy},
  \citenamefont {Bonn}, \citenamefont {Yoshida} \emph
  {et~al.}}]{comin2015symmetry}%
  \BibitemOpen
  \bibfield  {author} {\bibinfo {author} {\bibfnamefont {R.}~\bibnamefont
  {Comin}}, \bibinfo {author} {\bibfnamefont {R.}~\bibnamefont {Sutarto}},
  \bibinfo {author} {\bibfnamefont {F.}~\bibnamefont {He}}, \bibinfo {author}
  {\bibfnamefont {E.}~\bibnamefont {da~Silva~Neto}}, \bibinfo {author}
  {\bibfnamefont {L.}~\bibnamefont {Chauviere}}, \bibinfo {author}
  {\bibfnamefont {A.}~\bibnamefont {Fra{\~n}o}}, \bibinfo {author}
  {\bibfnamefont {R.}~\bibnamefont {Liang}}, \bibinfo {author} {\bibfnamefont
  {W.}~\bibnamefont {Hardy}}, \bibinfo {author} {\bibfnamefont
  {D.}~\bibnamefont {Bonn}}, \bibinfo {author} {\bibfnamefont {Y.}~\bibnamefont
  {Yoshida}},  \emph {et~al.},\ }\bibfield  {title} {\enquote {\bibinfo {title}
  {Symmetry of charge order in cuprates},}\ }\href@noop {} {\bibfield
  {journal} {\bibinfo  {journal} {Nature Materials}\ }\textbf {\bibinfo
  {volume} {14}},\ \bibinfo {pages} {796--800} (\bibinfo {year}
  {2015}{\natexlab{a}})}\BibitemShut {NoStop}%
\bibitem [{\citenamefont {Comin}\ \emph
  {et~al.}(2015{\natexlab{b}})\citenamefont {Comin}, \citenamefont {Sutarto},
  \citenamefont {da~Silva~Neto}, \citenamefont {Chauviere}, \citenamefont
  {Liang}, \citenamefont {Hardy}, \citenamefont {Bonn}, \citenamefont {He},
  \citenamefont {Sawatzky},\ and\ \citenamefont
  {Damascelli}}]{comin2015broken}%
  \BibitemOpen
  \bibfield  {author} {\bibinfo {author} {\bibfnamefont {R.}~\bibnamefont
  {Comin}}, \bibinfo {author} {\bibfnamefont {R.}~\bibnamefont {Sutarto}},
  \bibinfo {author} {\bibfnamefont {E.}~\bibnamefont {da~Silva~Neto}}, \bibinfo
  {author} {\bibfnamefont {L.}~\bibnamefont {Chauviere}}, \bibinfo {author}
  {\bibfnamefont {R.}~\bibnamefont {Liang}}, \bibinfo {author} {\bibfnamefont
  {W.}~\bibnamefont {Hardy}}, \bibinfo {author} {\bibfnamefont
  {D.}~\bibnamefont {Bonn}}, \bibinfo {author} {\bibfnamefont {F.}~\bibnamefont
  {He}}, \bibinfo {author} {\bibfnamefont {G.}~\bibnamefont {Sawatzky}}, \ and\
  \bibinfo {author} {\bibfnamefont {A.}~\bibnamefont {Damascelli}},\ }\bibfield
   {title} {\enquote {\bibinfo {title} {Broken translational and rotational
  symmetry via charge stripe order in underdoped \ce{YBa2Cu3O_{6+y}}},}\
  }\href@noop {} {\bibfield  {journal} {\bibinfo  {journal} {Science}\ }\textbf
  {\bibinfo {volume} {347}},\ \bibinfo {pages} {1335--1339} (\bibinfo {year}
  {2015}{\natexlab{b}})}\BibitemShut {NoStop}%
\bibitem [{\citenamefont {Comin}\ and\ \citenamefont
  {Damascelli}(2016)}]{comin2016resonant}%
  \BibitemOpen
  \bibfield  {author} {\bibinfo {author} {\bibfnamefont {R.}~\bibnamefont
  {Comin}}\ and\ \bibinfo {author} {\bibfnamefont {A.}~\bibnamefont
  {Damascelli}},\ }\bibfield  {title} {\enquote {\bibinfo {title} {Resonant
  x-ray scattering studies of charge order in cuprates},}\ }\href@noop {}
  {\bibfield  {journal} {\bibinfo  {journal} {Annual Review of Condensed Matter
  Physics}\ }\textbf {\bibinfo {volume} {7}},\ \bibinfo {pages} {369--405}
  (\bibinfo {year} {2016})}\BibitemShut {NoStop}%
\bibitem [{\citenamefont {Fradkin}\ and\ \citenamefont
  {Kivelson}(2012)}]{fradkin2012high}%
  \BibitemOpen
  \bibfield  {author} {\bibinfo {author} {\bibfnamefont {E.}~\bibnamefont
  {Fradkin}}\ and\ \bibinfo {author} {\bibfnamefont {S.~A.}\ \bibnamefont
  {Kivelson}},\ }\bibfield  {title} {\enquote {\bibinfo {title}
  {High-temperature superconductivity: Ineluctable complexity},}\ }\href@noop
  {} {\bibfield  {journal} {\bibinfo  {journal} {Nature Physics}\ }\textbf
  {\bibinfo {volume} {8}},\ \bibinfo {pages} {864--866} (\bibinfo {year}
  {2012})}\BibitemShut {NoStop}%
\bibitem [{\citenamefont {Fradkin}\ \emph {et~al.}(2015)\citenamefont
  {Fradkin}, \citenamefont {Kivelson},\ and\ \citenamefont
  {Tranquada}}]{fradkin2015colloquium}%
  \BibitemOpen
  \bibfield  {author} {\bibinfo {author} {\bibfnamefont {E.}~\bibnamefont
  {Fradkin}}, \bibinfo {author} {\bibfnamefont {S.~A.}\ \bibnamefont
  {Kivelson}}, \ and\ \bibinfo {author} {\bibfnamefont {J.~M.}\ \bibnamefont
  {Tranquada}},\ }\bibfield  {title} {\enquote {\bibinfo {title} {Colloquium:
  Theory of intertwined orders in high temperature superconductors},}\
  }\href@noop {} {\bibfield  {journal} {\bibinfo  {journal} {Reviews of Modern
  Physics}\ }\textbf {\bibinfo {volume} {87}},\ \bibinfo {pages} {457}
  (\bibinfo {year} {2015})}\BibitemShut {NoStop}%
\bibitem [{\citenamefont {Imry}\ and\ \citenamefont
  {Ma}(1975)}]{imry1975random}%
  \BibitemOpen
  \bibfield  {author} {\bibinfo {author} {\bibfnamefont {Y.}~\bibnamefont
  {Imry}}\ and\ \bibinfo {author} {\bibfnamefont {S.-K.}\ \bibnamefont {Ma}},\
  }\bibfield  {title} {\enquote {\bibinfo {title} {Random-field instability of
  the ordered state of continuous symmetry},}\ }\href@noop {} {\bibfield
  {journal} {\bibinfo  {journal} {Physical Review Letters}\ }\textbf {\bibinfo
  {volume} {35}},\ \bibinfo {pages} {1399} (\bibinfo {year}
  {1975})}\BibitemShut {NoStop}%
\bibitem [{\citenamefont {Nie}\ \emph {et~al.}(2014)\citenamefont {Nie},
  \citenamefont {Tarjus},\ and\ \citenamefont {Kivelson}}]{nie2014quenched}%
  \BibitemOpen
  \bibfield  {author} {\bibinfo {author} {\bibfnamefont {L.}~\bibnamefont
  {Nie}}, \bibinfo {author} {\bibfnamefont {G.}~\bibnamefont {Tarjus}}, \ and\
  \bibinfo {author} {\bibfnamefont {S.~A.}\ \bibnamefont {Kivelson}},\
  }\bibfield  {title} {\enquote {\bibinfo {title} {Quenched disorder and
  vestigial nematicity in the pseudogap regime of the cuprates},}\ }\href@noop
  {} {\bibfield  {journal} {\bibinfo  {journal} {Proceedings of the National
  Academy of Sciences}\ }\textbf {\bibinfo {volume} {111}},\ \bibinfo {pages}
  {7980--7985} (\bibinfo {year} {2014})}\BibitemShut {NoStop}%
\bibitem [{\citenamefont {da~Silva~Neto}\ \emph {et~al.}(2014)\citenamefont
  {da~Silva~Neto}, \citenamefont {Aynajian}, \citenamefont {Frano},
  \citenamefont {Comin}, \citenamefont {Schierle}, \citenamefont {Weschke},
  \citenamefont {Gyenis}, \citenamefont {Wen}, \citenamefont {Schneeloch},
  \citenamefont {Xu} \emph {et~al.}}]{da2014ubiquitous}%
  \BibitemOpen
  \bibfield  {author} {\bibinfo {author} {\bibfnamefont {E.~H.}\ \bibnamefont
  {da~Silva~Neto}}, \bibinfo {author} {\bibfnamefont {P.}~\bibnamefont
  {Aynajian}}, \bibinfo {author} {\bibfnamefont {A.}~\bibnamefont {Frano}},
  \bibinfo {author} {\bibfnamefont {R.}~\bibnamefont {Comin}}, \bibinfo
  {author} {\bibfnamefont {E.}~\bibnamefont {Schierle}}, \bibinfo {author}
  {\bibfnamefont {E.}~\bibnamefont {Weschke}}, \bibinfo {author} {\bibfnamefont
  {A.}~\bibnamefont {Gyenis}}, \bibinfo {author} {\bibfnamefont
  {J.}~\bibnamefont {Wen}}, \bibinfo {author} {\bibfnamefont {J.}~\bibnamefont
  {Schneeloch}}, \bibinfo {author} {\bibfnamefont {Z.}~\bibnamefont {Xu}},
  \emph {et~al.},\ }\bibfield  {title} {\enquote {\bibinfo {title} {Ubiquitous
  interplay between charge ordering and high-temperature superconductivity in
  cuprates},}\ }\href@noop {} {\bibfield  {journal} {\bibinfo  {journal}
  {Science}\ }\textbf {\bibinfo {volume} {343}},\ \bibinfo {pages} {393--396}
  (\bibinfo {year} {2014})}\BibitemShut {NoStop}%
\bibitem [{\citenamefont {Arguello}\ \emph {et~al.}(2015)\citenamefont
  {Arguello}, \citenamefont {Rosenthal}, \citenamefont {Andrade}, \citenamefont
  {Jin}, \citenamefont {Yeh}, \citenamefont {Zaki}, \citenamefont {Jia},
  \citenamefont {Cava}, \citenamefont {Fernandes}, \citenamefont {Millis} \emph
  {et~al.}}]{arguello2015quasiparticle}%
  \BibitemOpen
  \bibfield  {author} {\bibinfo {author} {\bibfnamefont {C.}~\bibnamefont
  {Arguello}}, \bibinfo {author} {\bibfnamefont {E.}~\bibnamefont {Rosenthal}},
  \bibinfo {author} {\bibfnamefont {E.}~\bibnamefont {Andrade}}, \bibinfo
  {author} {\bibfnamefont {W.}~\bibnamefont {Jin}}, \bibinfo {author}
  {\bibfnamefont {P.}~\bibnamefont {Yeh}}, \bibinfo {author} {\bibfnamefont
  {N.}~\bibnamefont {Zaki}}, \bibinfo {author} {\bibfnamefont {S.}~\bibnamefont
  {Jia}}, \bibinfo {author} {\bibfnamefont {R.}~\bibnamefont {Cava}}, \bibinfo
  {author} {\bibfnamefont {R.}~\bibnamefont {Fernandes}}, \bibinfo {author}
  {\bibfnamefont {A.}~\bibnamefont {Millis}},  \emph {et~al.},\ }\bibfield
  {title} {\enquote {\bibinfo {title} {Quasiparticle interference,
  quasiparticle interactions, and the origin of the charge density wave in
  {2H}-\ce{NbSe2}},}\ }\href@noop {} {\bibfield  {journal} {\bibinfo  {journal}
  {Physical Review Letters}\ }\textbf {\bibinfo {volume} {114}},\ \bibinfo
  {pages} {037001} (\bibinfo {year} {2015})}\BibitemShut {NoStop}%
\bibitem [{\citenamefont {El~Baggari}\ \emph {et~al.}(2018)\citenamefont
  {El~Baggari}, \citenamefont {Savitzky}, \citenamefont {Admasu}, \citenamefont
  {Kim}, \citenamefont {Cheong}, \citenamefont {Hovden},\ and\ \citenamefont
  {Kourkoutis}}]{el2018nature}%
  \BibitemOpen
  \bibfield  {author} {\bibinfo {author} {\bibfnamefont {I.}~\bibnamefont
  {El~Baggari}}, \bibinfo {author} {\bibfnamefont {B.~H.}\ \bibnamefont
  {Savitzky}}, \bibinfo {author} {\bibfnamefont {A.~S.}\ \bibnamefont
  {Admasu}}, \bibinfo {author} {\bibfnamefont {J.}~\bibnamefont {Kim}},
  \bibinfo {author} {\bibfnamefont {S.-W.}\ \bibnamefont {Cheong}}, \bibinfo
  {author} {\bibfnamefont {R.}~\bibnamefont {Hovden}}, \ and\ \bibinfo {author}
  {\bibfnamefont {L.~F.}\ \bibnamefont {Kourkoutis}},\ }\bibfield  {title}
  {\enquote {\bibinfo {title} {Nature and evolution of incommensurate charge
  order in manganites visualized with cryogenic scanning transmission electron
  microscopy},}\ }\href@noop {} {\bibfield  {journal} {\bibinfo  {journal}
  {Proceedings of the National Academy of Sciences}\ }\textbf {\bibinfo
  {volume} {115}},\ \bibinfo {pages} {1445--1450} (\bibinfo {year}
  {2018})}\BibitemShut {NoStop}%
\bibitem [{\citenamefont {Wu}\ \emph {et~al.}(2011)\citenamefont {Wu},
  \citenamefont {Mayaffre}, \citenamefont {Kr{\"a}mer}, \citenamefont
  {Horvati{\'c}}, \citenamefont {Berthier}, \citenamefont {Hardy},
  \citenamefont {Liang}, \citenamefont {Bonn},\ and\ \citenamefont
  {Julien}}]{wu2011magnetic}%
  \BibitemOpen
  \bibfield  {author} {\bibinfo {author} {\bibfnamefont {T.}~\bibnamefont
  {Wu}}, \bibinfo {author} {\bibfnamefont {H.}~\bibnamefont {Mayaffre}},
  \bibinfo {author} {\bibfnamefont {S.}~\bibnamefont {Kr{\"a}mer}}, \bibinfo
  {author} {\bibfnamefont {M.}~\bibnamefont {Horvati{\'c}}}, \bibinfo {author}
  {\bibfnamefont {C.}~\bibnamefont {Berthier}}, \bibinfo {author}
  {\bibfnamefont {W.}~\bibnamefont {Hardy}}, \bibinfo {author} {\bibfnamefont
  {R.}~\bibnamefont {Liang}}, \bibinfo {author} {\bibfnamefont
  {D.}~\bibnamefont {Bonn}}, \ and\ \bibinfo {author} {\bibfnamefont {M.-H.}\
  \bibnamefont {Julien}},\ }\bibfield  {title} {\enquote {\bibinfo {title}
  {Magnetic-field-induced charge-stripe order in the high-temperature
  superconductor \ce{YBa2Cu3Oy}},}\ }\href@noop {} {\bibfield  {journal}
  {\bibinfo  {journal} {Nature}\ }\textbf {\bibinfo {volume} {477}},\ \bibinfo
  {pages} {191--194} (\bibinfo {year} {2011})}\BibitemShut {NoStop}%
\bibitem [{\citenamefont {Wu}\ \emph {et~al.}(2015)\citenamefont {Wu},
  \citenamefont {Mayaffre}, \citenamefont {Kr{\"a}mer}, \citenamefont
  {Horvati{\'c}}, \citenamefont {Berthier}, \citenamefont {Hardy},
  \citenamefont {Liang}, \citenamefont {Bonn},\ and\ \citenamefont
  {Julien}}]{wu2015incipient}%
  \BibitemOpen
  \bibfield  {author} {\bibinfo {author} {\bibfnamefont {T.}~\bibnamefont
  {Wu}}, \bibinfo {author} {\bibfnamefont {H.}~\bibnamefont {Mayaffre}},
  \bibinfo {author} {\bibfnamefont {S.}~\bibnamefont {Kr{\"a}mer}}, \bibinfo
  {author} {\bibfnamefont {M.}~\bibnamefont {Horvati{\'c}}}, \bibinfo {author}
  {\bibfnamefont {C.}~\bibnamefont {Berthier}}, \bibinfo {author}
  {\bibfnamefont {W.}~\bibnamefont {Hardy}}, \bibinfo {author} {\bibfnamefont
  {R.}~\bibnamefont {Liang}}, \bibinfo {author} {\bibfnamefont
  {D.}~\bibnamefont {Bonn}}, \ and\ \bibinfo {author} {\bibfnamefont {M.-H.}\
  \bibnamefont {Julien}},\ }\bibfield  {title} {\enquote {\bibinfo {title}
  {Incipient charge order observed by {NMR} in the normal state of
  \ce{YBa2Cu3Oy}},}\ }\href@noop {} {\bibfield  {journal} {\bibinfo  {journal}
  {Nature Communications}\ }\textbf {\bibinfo {volume} {6}},\ \bibinfo {pages}
  {6438} (\bibinfo {year} {2015})}\BibitemShut {NoStop}%
\bibitem [{\citenamefont {Kim}\ \emph {et~al.}(2008)\citenamefont {Kim},
  \citenamefont {Jin}, \citenamefont {Moon}, \citenamefont {Kim}, \citenamefont
  {Park}, \citenamefont {Leem}, \citenamefont {Yu}, \citenamefont {Noh},
  \citenamefont {Kim}, \citenamefont {Oh} \emph {et~al.}}]{kim2008novel}%
  \BibitemOpen
  \bibfield  {author} {\bibinfo {author} {\bibfnamefont {B.}~\bibnamefont
  {Kim}}, \bibinfo {author} {\bibfnamefont {H.}~\bibnamefont {Jin}}, \bibinfo
  {author} {\bibfnamefont {S.}~\bibnamefont {Moon}}, \bibinfo {author}
  {\bibfnamefont {J.-Y.}\ \bibnamefont {Kim}}, \bibinfo {author} {\bibfnamefont
  {B.-G.}\ \bibnamefont {Park}}, \bibinfo {author} {\bibfnamefont
  {C.}~\bibnamefont {Leem}}, \bibinfo {author} {\bibfnamefont {J.}~\bibnamefont
  {Yu}}, \bibinfo {author} {\bibfnamefont {T.}~\bibnamefont {Noh}}, \bibinfo
  {author} {\bibfnamefont {C.}~\bibnamefont {Kim}}, \bibinfo {author}
  {\bibfnamefont {S.-J.}\ \bibnamefont {Oh}},  \emph {et~al.},\ }\bibfield
  {title} {\enquote {\bibinfo {title} {Novel ${J}_{eff} = 1/2$ {Mott} state
  induced by relativistic spin-orbit coupling in \ce{Sr2IrO4}},}\ }\href@noop
  {} {\bibfield  {journal} {\bibinfo  {journal} {Physical Review Letters}\
  }\textbf {\bibinfo {volume} {101}},\ \bibinfo {pages} {076402} (\bibinfo
  {year} {2008})}\BibitemShut {NoStop}%
\bibitem [{\citenamefont {Kim}\ \emph {et~al.}(2009)\citenamefont {Kim},
  \citenamefont {Ohsumi}, \citenamefont {Komesu}, \citenamefont {Sakai},
  \citenamefont {Morita}, \citenamefont {Takagi},\ and\ \citenamefont
  {Arima}}]{kim2009phase}%
  \BibitemOpen
  \bibfield  {author} {\bibinfo {author} {\bibfnamefont {B.}~\bibnamefont
  {Kim}}, \bibinfo {author} {\bibfnamefont {H.}~\bibnamefont {Ohsumi}},
  \bibinfo {author} {\bibfnamefont {T.}~\bibnamefont {Komesu}}, \bibinfo
  {author} {\bibfnamefont {S.}~\bibnamefont {Sakai}}, \bibinfo {author}
  {\bibfnamefont {T.}~\bibnamefont {Morita}}, \bibinfo {author} {\bibfnamefont
  {H.}~\bibnamefont {Takagi}}, \ and\ \bibinfo {author} {\bibfnamefont
  {T.}~\bibnamefont {Arima}},\ }\bibfield  {title} {\enquote {\bibinfo {title}
  {Phase-sensitive observation of a spin-orbital {Mott} state in
  \ce{Sr2IrO4}},}\ }\href@noop {} {\bibfield  {journal} {\bibinfo  {journal}
  {Science}\ }\textbf {\bibinfo {volume} {323}},\ \bibinfo {pages} {1329--1332}
  (\bibinfo {year} {2009})}\BibitemShut {NoStop}%
\bibitem [{\citenamefont {Kim}\ \emph {et~al.}(2014)\citenamefont {Kim},
  \citenamefont {Krupin}, \citenamefont {Denlinger}, \citenamefont {Bostwick},
  \citenamefont {Rotenberg}, \citenamefont {Zhao}, \citenamefont {Mitchell},
  \citenamefont {Allen},\ and\ \citenamefont {Kim}}]{kim2014fermi}%
  \BibitemOpen
  \bibfield  {author} {\bibinfo {author} {\bibfnamefont {Y.}~\bibnamefont
  {Kim}}, \bibinfo {author} {\bibfnamefont {O.}~\bibnamefont {Krupin}},
  \bibinfo {author} {\bibfnamefont {J.}~\bibnamefont {Denlinger}}, \bibinfo
  {author} {\bibfnamefont {A.}~\bibnamefont {Bostwick}}, \bibinfo {author}
  {\bibfnamefont {E.}~\bibnamefont {Rotenberg}}, \bibinfo {author}
  {\bibfnamefont {Q.}~\bibnamefont {Zhao}}, \bibinfo {author} {\bibfnamefont
  {J.}~\bibnamefont {Mitchell}}, \bibinfo {author} {\bibfnamefont
  {J.}~\bibnamefont {Allen}}, \ and\ \bibinfo {author} {\bibfnamefont
  {B.}~\bibnamefont {Kim}},\ }\bibfield  {title} {\enquote {\bibinfo {title}
  {Fermi arcs in a doped pseudospin-{1/2 Heisenberg} antiferromagnet},}\
  }\href@noop {} {\bibfield  {journal} {\bibinfo  {journal} {Science}\ }\textbf
  {\bibinfo {volume} {345}},\ \bibinfo {pages} {187--190} (\bibinfo {year}
  {2014})}\BibitemShut {NoStop}%
\bibitem [{\citenamefont {De~La~Torre}\ \emph {et~al.}(2015)\citenamefont
  {De~La~Torre}, \citenamefont {Walker}, \citenamefont {Bruno}, \citenamefont
  {Ricc{\'o}}, \citenamefont {Wang}, \citenamefont {Lezama}, \citenamefont
  {Scheerer}, \citenamefont {Giriat}, \citenamefont {Jaccard}, \citenamefont
  {Berthod} \emph {et~al.}}]{de2015collapse}%
  \BibitemOpen
  \bibfield  {author} {\bibinfo {author} {\bibfnamefont {A.}~\bibnamefont
  {De~La~Torre}}, \bibinfo {author} {\bibfnamefont {S.~M.}\ \bibnamefont
  {Walker}}, \bibinfo {author} {\bibfnamefont {F.~Y.}\ \bibnamefont {Bruno}},
  \bibinfo {author} {\bibfnamefont {S.}~\bibnamefont {Ricc{\'o}}}, \bibinfo
  {author} {\bibfnamefont {Z.}~\bibnamefont {Wang}}, \bibinfo {author}
  {\bibfnamefont {I.~G.}\ \bibnamefont {Lezama}}, \bibinfo {author}
  {\bibfnamefont {G.}~\bibnamefont {Scheerer}}, \bibinfo {author}
  {\bibfnamefont {G.}~\bibnamefont {Giriat}}, \bibinfo {author} {\bibfnamefont
  {D.}~\bibnamefont {Jaccard}}, \bibinfo {author} {\bibfnamefont
  {C.}~\bibnamefont {Berthod}},  \emph {et~al.},\ }\bibfield  {title} {\enquote
  {\bibinfo {title} {Collapse of the {Mott} gap and emergence of a nodal liquid
  in lightly doped \ce{Sr2IrO4}},}\ }\href@noop {} {\bibfield  {journal}
  {\bibinfo  {journal} {Physical Review Letters}\ }\textbf {\bibinfo {volume}
  {115}},\ \bibinfo {pages} {176402} (\bibinfo {year} {2015})}\BibitemShut
  {NoStop}%
\bibitem [{\citenamefont {Cao}\ \emph {et~al.}(2016)\citenamefont {Cao},
  \citenamefont {Wang}, \citenamefont {Waugh}, \citenamefont {Reber},
  \citenamefont {Li}, \citenamefont {Zhou}, \citenamefont {Parham},
  \citenamefont {Park}, \citenamefont {Plumb}, \citenamefont {Rotenberg} \emph
  {et~al.}}]{cao2016hallmarks}%
  \BibitemOpen
  \bibfield  {author} {\bibinfo {author} {\bibfnamefont {Y.}~\bibnamefont
  {Cao}}, \bibinfo {author} {\bibfnamefont {Q.}~\bibnamefont {Wang}}, \bibinfo
  {author} {\bibfnamefont {J.~A.}\ \bibnamefont {Waugh}}, \bibinfo {author}
  {\bibfnamefont {T.~J.}\ \bibnamefont {Reber}}, \bibinfo {author}
  {\bibfnamefont {H.}~\bibnamefont {Li}}, \bibinfo {author} {\bibfnamefont
  {X.}~\bibnamefont {Zhou}}, \bibinfo {author} {\bibfnamefont {S.}~\bibnamefont
  {Parham}}, \bibinfo {author} {\bibfnamefont {S.-R.}\ \bibnamefont {Park}},
  \bibinfo {author} {\bibfnamefont {N.~C.}\ \bibnamefont {Plumb}}, \bibinfo
  {author} {\bibfnamefont {E.}~\bibnamefont {Rotenberg}},  \emph {et~al.},\
  }\bibfield  {title} {\enquote {\bibinfo {title} {Hallmarks of the
  {Mott}-metal crossover in the hole-doped pseudospin-1/2 {Mott} insulator
  \ce{Sr2IrO4}},}\ }\href@noop {} {\bibfield  {journal} {\bibinfo  {journal}
  {Nature communications}\ }\textbf {\bibinfo {volume} {7}},\ \bibinfo {pages}
  {11367} (\bibinfo {year} {2016})}\BibitemShut {NoStop}%
\bibitem [{\citenamefont {Battisti}\ \emph {et~al.}(2017)\citenamefont
  {Battisti}, \citenamefont {Bastiaans}, \citenamefont {Fedoseev},
  \citenamefont {De~La~Torre}, \citenamefont {Iliopoulos}, \citenamefont
  {Tamai}, \citenamefont {Hunter}, \citenamefont {Perry}, \citenamefont
  {Zaanen}, \citenamefont {Baumberger} \emph
  {et~al.}}]{battisti2017universality}%
  \BibitemOpen
  \bibfield  {author} {\bibinfo {author} {\bibfnamefont {I.}~\bibnamefont
  {Battisti}}, \bibinfo {author} {\bibfnamefont {K.~M.}\ \bibnamefont
  {Bastiaans}}, \bibinfo {author} {\bibfnamefont {V.}~\bibnamefont {Fedoseev}},
  \bibinfo {author} {\bibfnamefont {A.}~\bibnamefont {De~La~Torre}}, \bibinfo
  {author} {\bibfnamefont {N.}~\bibnamefont {Iliopoulos}}, \bibinfo {author}
  {\bibfnamefont {A.}~\bibnamefont {Tamai}}, \bibinfo {author} {\bibfnamefont
  {E.~C.}\ \bibnamefont {Hunter}}, \bibinfo {author} {\bibfnamefont {R.~S.}\
  \bibnamefont {Perry}}, \bibinfo {author} {\bibfnamefont {J.}~\bibnamefont
  {Zaanen}}, \bibinfo {author} {\bibfnamefont {F.}~\bibnamefont {Baumberger}},
  \emph {et~al.},\ }\bibfield  {title} {\enquote {\bibinfo {title}
  {Universality of pseudogap and emergent order in lightly doped {Mott}
  insulators},}\ }\href@noop {} {\bibfield  {journal} {\bibinfo  {journal}
  {Nature Physics}\ }\textbf {\bibinfo {volume} {13}},\ \bibinfo {pages}
  {21--25} (\bibinfo {year} {2017})}\BibitemShut {NoStop}%
\bibitem [{\citenamefont {Zhao}\ \emph {et~al.}(2016)\citenamefont {Zhao},
  \citenamefont {Torchinsky}, \citenamefont {Chu}, \citenamefont {Ivanov},
  \citenamefont {Lifshitz}, \citenamefont {Flint}, \citenamefont {Qi},
  \citenamefont {Cao},\ and\ \citenamefont {Hsieh}}]{zhao2016evidence}%
  \BibitemOpen
  \bibfield  {author} {\bibinfo {author} {\bibfnamefont {L.}~\bibnamefont
  {Zhao}}, \bibinfo {author} {\bibfnamefont {D.}~\bibnamefont {Torchinsky}},
  \bibinfo {author} {\bibfnamefont {H.}~\bibnamefont {Chu}}, \bibinfo {author}
  {\bibfnamefont {V.}~\bibnamefont {Ivanov}}, \bibinfo {author} {\bibfnamefont
  {R.}~\bibnamefont {Lifshitz}}, \bibinfo {author} {\bibfnamefont
  {R.}~\bibnamefont {Flint}}, \bibinfo {author} {\bibfnamefont
  {T.}~\bibnamefont {Qi}}, \bibinfo {author} {\bibfnamefont {G.}~\bibnamefont
  {Cao}}, \ and\ \bibinfo {author} {\bibfnamefont {D.}~\bibnamefont {Hsieh}},\
  }\bibfield  {title} {\enquote {\bibinfo {title} {Evidence of an odd-parity
  hidden order in a spin-orbit coupled correlated iridate},}\ }\href@noop {}
  {\bibfield  {journal} {\bibinfo  {journal} {Nature Physics}\ }\textbf
  {\bibinfo {volume} {12}},\ \bibinfo {pages} {32--36} (\bibinfo {year}
  {2016})}\BibitemShut {NoStop}%
\bibitem [{\citenamefont {Jeong}\ \emph {et~al.}(2017)\citenamefont {Jeong},
  \citenamefont {Sidis}, \citenamefont {Louat}, \citenamefont {Brouet},\ and\
  \citenamefont {Bourges}}]{jeong2017time}%
  \BibitemOpen
  \bibfield  {author} {\bibinfo {author} {\bibfnamefont {J.}~\bibnamefont
  {Jeong}}, \bibinfo {author} {\bibfnamefont {Y.}~\bibnamefont {Sidis}},
  \bibinfo {author} {\bibfnamefont {A.}~\bibnamefont {Louat}}, \bibinfo
  {author} {\bibfnamefont {V.}~\bibnamefont {Brouet}}, \ and\ \bibinfo {author}
  {\bibfnamefont {P.}~\bibnamefont {Bourges}},\ }\bibfield  {title} {\enquote
  {\bibinfo {title} {Time-reversal symmetry breaking hidden order in
  ${Sr_2(Ir,Rh)O_4}$},}\ }\href@noop {} {\bibfield  {journal} {\bibinfo
  {journal} {Nature Communications}\ }\textbf {\bibinfo {volume} {8}},\
  \bibinfo {pages} {15119} (\bibinfo {year} {2017})}\BibitemShut {NoStop}%
\bibitem [{\citenamefont {Yan}\ \emph {et~al.}(2015)\citenamefont {Yan},
  \citenamefont {Ren}, \citenamefont {Xu}, \citenamefont {Xie}, \citenamefont
  {Tao}, \citenamefont {Choi}, \citenamefont {Lee}, \citenamefont {Choi},
  \citenamefont {Zhang},\ and\ \citenamefont {Feng}}]{yan2015electron}%
  \BibitemOpen
  \bibfield  {author} {\bibinfo {author} {\bibfnamefont {Y.}~\bibnamefont
  {Yan}}, \bibinfo {author} {\bibfnamefont {M.}~\bibnamefont {Ren}}, \bibinfo
  {author} {\bibfnamefont {H.}~\bibnamefont {Xu}}, \bibinfo {author}
  {\bibfnamefont {B.}~\bibnamefont {Xie}}, \bibinfo {author} {\bibfnamefont
  {R.}~\bibnamefont {Tao}}, \bibinfo {author} {\bibfnamefont {H.}~\bibnamefont
  {Choi}}, \bibinfo {author} {\bibfnamefont {N.}~\bibnamefont {Lee}}, \bibinfo
  {author} {\bibfnamefont {Y.}~\bibnamefont {Choi}}, \bibinfo {author}
  {\bibfnamefont {T.}~\bibnamefont {Zhang}}, \ and\ \bibinfo {author}
  {\bibfnamefont {D.}~\bibnamefont {Feng}},\ }\bibfield  {title} {\enquote
  {\bibinfo {title} {Electron-doped \ce{Sr2IrO4}: an analogue of hole-doped
  cuprate superconductors demonstrated by scanning tunneling microscopy},}\
  }\href@noop {} {\bibfield  {journal} {\bibinfo  {journal} {Physical Review
  X}\ }\textbf {\bibinfo {volume} {5}},\ \bibinfo {pages} {041018} (\bibinfo
  {year} {2015})}\BibitemShut {NoStop}%
\bibitem [{\citenamefont {Kim}\ \emph {et~al.}(2016)\citenamefont {Kim},
  \citenamefont {Sung}, \citenamefont {Denlinger},\ and\ \citenamefont
  {Kim}}]{kim2016observation}%
  \BibitemOpen
  \bibfield  {author} {\bibinfo {author} {\bibfnamefont {Y.}~\bibnamefont
  {Kim}}, \bibinfo {author} {\bibfnamefont {N.}~\bibnamefont {Sung}}, \bibinfo
  {author} {\bibfnamefont {J.}~\bibnamefont {Denlinger}}, \ and\ \bibinfo
  {author} {\bibfnamefont {B.}~\bibnamefont {Kim}},\ }\bibfield  {title}
  {\enquote {\bibinfo {title} {Observation of a \textit{d}-wave gap in
  electron-doped \ce{Sr2IrO4}},}\ }\href@noop {} {\bibfield  {journal}
  {\bibinfo  {journal} {Nature Physics}\ }\textbf {\bibinfo {volume} {12}},\
  \bibinfo {pages} {37--41} (\bibinfo {year} {2016})}\BibitemShut {NoStop}%
\bibitem [{\citenamefont {Li}\ \emph {et~al.}(2013)\citenamefont {Li},
  \citenamefont {Kong}, \citenamefont {Qi}, \citenamefont {Jin}, \citenamefont
  {Yuan}, \citenamefont {DeLong}, \citenamefont {Schlottmann},\ and\
  \citenamefont {Cao}}]{li2013tuning}%
  \BibitemOpen
  \bibfield  {author} {\bibinfo {author} {\bibfnamefont {L.}~\bibnamefont
  {Li}}, \bibinfo {author} {\bibfnamefont {P.}~\bibnamefont {Kong}}, \bibinfo
  {author} {\bibfnamefont {T.}~\bibnamefont {Qi}}, \bibinfo {author}
  {\bibfnamefont {C.}~\bibnamefont {Jin}}, \bibinfo {author} {\bibfnamefont
  {S.}~\bibnamefont {Yuan}}, \bibinfo {author} {\bibfnamefont {L.}~\bibnamefont
  {DeLong}}, \bibinfo {author} {\bibfnamefont {P.}~\bibnamefont {Schlottmann}},
  \ and\ \bibinfo {author} {\bibfnamefont {G.}~\bibnamefont {Cao}},\ }\bibfield
   {title} {\enquote {\bibinfo {title} {Tuning the ${J}_{eff} = 1/2$ insulating
  state via electron doping and pressure in the double-layered iridate
  \ce{Sr3Ir2O7}},}\ }\href@noop {} {\bibfield  {journal} {\bibinfo  {journal}
  {Physical Review B}\ }\textbf {\bibinfo {volume} {87}},\ \bibinfo {pages}
  {235127} (\bibinfo {year} {2013})}\BibitemShut {NoStop}%
\bibitem [{\citenamefont {Hogan}\ \emph {et~al.}(2015)\citenamefont {Hogan},
  \citenamefont {Yamani}, \citenamefont {Walkup}, \citenamefont {Chen},
  \citenamefont {Dally}, \citenamefont {Ward}, \citenamefont {Dean},
  \citenamefont {Hill}, \citenamefont {Islam}, \citenamefont {Madhavan} \emph
  {et~al.}}]{hogan2015first}%
  \BibitemOpen
  \bibfield  {author} {\bibinfo {author} {\bibfnamefont {T.}~\bibnamefont
  {Hogan}}, \bibinfo {author} {\bibfnamefont {Z.}~\bibnamefont {Yamani}},
  \bibinfo {author} {\bibfnamefont {D.}~\bibnamefont {Walkup}}, \bibinfo
  {author} {\bibfnamefont {X.}~\bibnamefont {Chen}}, \bibinfo {author}
  {\bibfnamefont {R.}~\bibnamefont {Dally}}, \bibinfo {author} {\bibfnamefont
  {T.~Z.}\ \bibnamefont {Ward}}, \bibinfo {author} {\bibfnamefont
  {M.}~\bibnamefont {Dean}}, \bibinfo {author} {\bibfnamefont {J.}~\bibnamefont
  {Hill}}, \bibinfo {author} {\bibfnamefont {Z.}~\bibnamefont {Islam}},
  \bibinfo {author} {\bibfnamefont {V.}~\bibnamefont {Madhavan}},  \emph
  {et~al.},\ }\bibfield  {title} {\enquote {\bibinfo {title} {First-order
  melting of a weak spin-orbit {Mott} insulator into a correlated metal},}\
  }\href@noop {} {\bibfield  {journal} {\bibinfo  {journal} {Physical Review
  Letters}\ }\textbf {\bibinfo {volume} {114}},\ \bibinfo {pages} {257203}
  (\bibinfo {year} {2015})}\BibitemShut {NoStop}%
\bibitem [{\citenamefont {Chu}\ \emph {et~al.}(2017)\citenamefont {Chu},
  \citenamefont {Zhao}, \citenamefont {de~la Torre}, \citenamefont {Hogan},
  \citenamefont {Wilson},\ and\ \citenamefont {Hsieh}}]{chu2017charge}%
  \BibitemOpen
  \bibfield  {author} {\bibinfo {author} {\bibfnamefont {H.}~\bibnamefont
  {Chu}}, \bibinfo {author} {\bibfnamefont {L.}~\bibnamefont {Zhao}}, \bibinfo
  {author} {\bibfnamefont {A.}~\bibnamefont {de~la Torre}}, \bibinfo {author}
  {\bibfnamefont {T.}~\bibnamefont {Hogan}}, \bibinfo {author} {\bibfnamefont
  {S.}~\bibnamefont {Wilson}}, \ and\ \bibinfo {author} {\bibfnamefont
  {D.}~\bibnamefont {Hsieh}},\ }\bibfield  {title} {\enquote {\bibinfo {title}
  {A charge density wave-like instability in a doped spin-orbit-assisted weak
  {Mott} insulator},}\ }\href@noop {} {\bibfield  {journal} {\bibinfo
  {journal} {Nature Materials}\ }\textbf {\bibinfo {volume} {16}},\ \bibinfo
  {pages} {200--203} (\bibinfo {year} {2017})}\BibitemShut {NoStop}%
\bibitem [{\citenamefont {Hogan}\ \emph {et~al.}(2016)\citenamefont {Hogan},
  \citenamefont {Bjaalie}, \citenamefont {Zhao}, \citenamefont {Belvin},
  \citenamefont {Wang}, \citenamefont {Van~de Walle}, \citenamefont {Hsieh},\
  and\ \citenamefont {Wilson}}]{hogan2016structural}%
  \BibitemOpen
  \bibfield  {author} {\bibinfo {author} {\bibfnamefont {T.}~\bibnamefont
  {Hogan}}, \bibinfo {author} {\bibfnamefont {L.}~\bibnamefont {Bjaalie}},
  \bibinfo {author} {\bibfnamefont {L.}~\bibnamefont {Zhao}}, \bibinfo {author}
  {\bibfnamefont {C.}~\bibnamefont {Belvin}}, \bibinfo {author} {\bibfnamefont
  {X.}~\bibnamefont {Wang}}, \bibinfo {author} {\bibfnamefont {C.~G.}\
  \bibnamefont {Van~de Walle}}, \bibinfo {author} {\bibfnamefont
  {D.}~\bibnamefont {Hsieh}}, \ and\ \bibinfo {author} {\bibfnamefont {S.~D.}\
  \bibnamefont {Wilson}},\ }\bibfield  {title} {\enquote {\bibinfo {title}
  {Structural investigation of the bilayer iridate \ce{Sr3Ir2O7}},}\
  }\href@noop {} {\bibfield  {journal} {\bibinfo  {journal} {Physical Review
  B}\ }\textbf {\bibinfo {volume} {93}},\ \bibinfo {pages} {134110} (\bibinfo
  {year} {2016})}\BibitemShut {NoStop}%
\bibitem [{\citenamefont {Hogan}\ \emph {et~al.}(2017)\citenamefont {Hogan},
  \citenamefont {Wang}, \citenamefont {Chu}, \citenamefont {Hsieh},\ and\
  \citenamefont {Wilson}}]{hogan2017doping}%
  \BibitemOpen
  \bibfield  {author} {\bibinfo {author} {\bibfnamefont {T.}~\bibnamefont
  {Hogan}}, \bibinfo {author} {\bibfnamefont {X.}~\bibnamefont {Wang}},
  \bibinfo {author} {\bibfnamefont {H.}~\bibnamefont {Chu}}, \bibinfo {author}
  {\bibfnamefont {D.}~\bibnamefont {Hsieh}}, \ and\ \bibinfo {author}
  {\bibfnamefont {S.~D.}\ \bibnamefont {Wilson}},\ }\bibfield  {title}
  {\enquote {\bibinfo {title} {Doping-driven structural distortion in the
  bilayer iridate \ce{(Sr_{1-x}La_{x})_{3}Ir2O7}},}\ }\href@noop {} {\bibfield
  {journal} {\bibinfo  {journal} {Physical Review B}\ }\textbf {\bibinfo
  {volume} {95}},\ \bibinfo {pages} {174109} (\bibinfo {year}
  {2017})}\BibitemShut {NoStop}%
\bibitem [{\citenamefont {Gretarsson}\ \emph {et~al.}(2016)\citenamefont
  {Gretarsson}, \citenamefont {Sung}, \citenamefont {H{\"o}ppner},
  \citenamefont {Kim}, \citenamefont {Keimer},\ and\ \citenamefont
  {Le~Tacon}}]{gretarsson2016two}%
  \BibitemOpen
  \bibfield  {author} {\bibinfo {author} {\bibfnamefont {H.}~\bibnamefont
  {Gretarsson}}, \bibinfo {author} {\bibfnamefont {N.}~\bibnamefont {Sung}},
  \bibinfo {author} {\bibfnamefont {M.}~\bibnamefont {H{\"o}ppner}}, \bibinfo
  {author} {\bibfnamefont {B.}~\bibnamefont {Kim}}, \bibinfo {author}
  {\bibfnamefont {B.}~\bibnamefont {Keimer}}, \ and\ \bibinfo {author}
  {\bibfnamefont {M.}~\bibnamefont {Le~Tacon}},\ }\bibfield  {title} {\enquote
  {\bibinfo {title} {Two-magnon {Raman} scattering and pseudospin-lattice
  interactions in \ce{Sr2IrO4} and \ce{Sr3Ir2O7}},}\ }\href@noop {} {\bibfield
  {journal} {\bibinfo  {journal} {Physical Review Letters}\ }\textbf {\bibinfo
  {volume} {116}},\ \bibinfo {pages} {136401} (\bibinfo {year}
  {2016})}\BibitemShut {NoStop}%
\bibitem [{Sup()}]{Supplementary}%
  \BibitemOpen
  \href@noop {} {\ }\bibinfo {note} {See Supplementary Material for Raman
  tensor analysis and selection rules of Raman modes at room
  temperature.}\BibitemShut {Stop}%
\bibitem [{\citenamefont {Burns}\ and\ \citenamefont
  {Scott}(1970)}]{burns1970raman}%
  \BibitemOpen
  \bibfield  {author} {\bibinfo {author} {\bibfnamefont {G.}~\bibnamefont
  {Burns}}\ and\ \bibinfo {author} {\bibfnamefont {B.~A.}\ \bibnamefont
  {Scott}},\ }\bibfield  {title} {\enquote {\bibinfo {title} {Raman studies of
  underdamped soft modes in \ce{PbTiO3}},}\ }\href@noop {} {\bibfield
  {journal} {\bibinfo  {journal} {Physical Review Letters}\ }\textbf {\bibinfo
  {volume} {25}},\ \bibinfo {pages} {167} (\bibinfo {year} {1970})}\BibitemShut
  {NoStop}%
\bibitem [{\citenamefont {Travaglini}\ \emph {et~al.}(1983)\citenamefont
  {Travaglini}, \citenamefont {M{\"o}rke},\ and\ \citenamefont
  {Wachter}}]{travaglini1983cdw}%
  \BibitemOpen
  \bibfield  {author} {\bibinfo {author} {\bibfnamefont {G.}~\bibnamefont
  {Travaglini}}, \bibinfo {author} {\bibfnamefont {I.}~\bibnamefont
  {M{\"o}rke}}, \ and\ \bibinfo {author} {\bibfnamefont {P.}~\bibnamefont
  {Wachter}},\ }\bibfield  {title} {\enquote {\bibinfo {title} {{CDW} evidence
  in one-dimensional \ce{K_{0.3}MoO3} by means of {Raman} scattering},}\
  }\href@noop {} {\bibfield  {journal} {\bibinfo  {journal} {Solid State
  Communications}\ }\textbf {\bibinfo {volume} {45}},\ \bibinfo {pages}
  {289--292} (\bibinfo {year} {1983})}\BibitemShut {NoStop}%
\bibitem [{\citenamefont {Tsang}\ \emph {et~al.}(1976)\citenamefont {Tsang},
  \citenamefont {Smith~Jr},\ and\ \citenamefont {Shafer}}]{tsang1976raman}%
  \BibitemOpen
  \bibfield  {author} {\bibinfo {author} {\bibfnamefont {J.}~\bibnamefont
  {Tsang}}, \bibinfo {author} {\bibfnamefont {J.}~\bibnamefont {Smith~Jr}}, \
  and\ \bibinfo {author} {\bibfnamefont {M.}~\bibnamefont {Shafer}},\
  }\bibfield  {title} {\enquote {\bibinfo {title} {Raman spectroscopy of soft
  modes at the charge-density-wave phase transition in {2H}-\ce{NbSe2}},}\
  }\href@noop {} {\bibfield  {journal} {\bibinfo  {journal} {Physical Review
  Letters}\ }\textbf {\bibinfo {volume} {37}},\ \bibinfo {pages} {1407}
  (\bibinfo {year} {1976})}\BibitemShut {NoStop}%
\bibitem [{\citenamefont {Gor'kov}\ and\ \citenamefont
  {Gr{\"u}ner}(1989)}]{gor2012charge}%
  \BibitemOpen
  \bibfield  {author} {\bibinfo {author} {\bibfnamefont {L.~P.}\ \bibnamefont
  {Gor'kov}}\ and\ \bibinfo {author} {\bibfnamefont {G.}~\bibnamefont
  {Gr{\"u}ner}},\ }\href@noop {} {\emph {\bibinfo {title} {Charge density waves
  in solids}}}\ (\bibinfo  {publisher} {Elsevier, New York},\ \bibinfo {year}
  {1989})\ \bibinfo {note} {{Vol}. 25, p. 494}\BibitemShut {NoStop}%
\bibitem [{\citenamefont {Torchinsky}\ \emph {et~al.}(2013)\citenamefont
  {Torchinsky}, \citenamefont {Mahmood}, \citenamefont {Bollinger},
  \citenamefont {Bo{\v{z}}ovi{\'c}},\ and\ \citenamefont
  {Gedik}}]{torchinsky2013fluctuating}%
  \BibitemOpen
  \bibfield  {author} {\bibinfo {author} {\bibfnamefont {D.~H.}\ \bibnamefont
  {Torchinsky}}, \bibinfo {author} {\bibfnamefont {F.}~\bibnamefont {Mahmood}},
  \bibinfo {author} {\bibfnamefont {A.~T.}\ \bibnamefont {Bollinger}}, \bibinfo
  {author} {\bibfnamefont {I.}~\bibnamefont {Bo{\v{z}}ovi{\'c}}}, \ and\
  \bibinfo {author} {\bibfnamefont {N.}~\bibnamefont {Gedik}},\ }\bibfield
  {title} {\enquote {\bibinfo {title} {Fluctuating charge-density waves in a
  cuprate superconductor},}\ }\href@noop {} {\bibfield  {journal} {\bibinfo
  {journal} {Nature Materials}\ }\textbf {\bibinfo {volume} {12}},\ \bibinfo
  {pages} {387--391} (\bibinfo {year} {2013})}\BibitemShut {NoStop}%
\bibitem [{\citenamefont {Hinton}\ \emph {et~al.}(2013)\citenamefont {Hinton},
  \citenamefont {Koralek}, \citenamefont {Lu}, \citenamefont {Vishwanath},
  \citenamefont {Orenstein}, \citenamefont {Bonn}, \citenamefont {Hardy},\ and\
  \citenamefont {Liang}}]{hinton2013new}%
  \BibitemOpen
  \bibfield  {author} {\bibinfo {author} {\bibfnamefont {J.}~\bibnamefont
  {Hinton}}, \bibinfo {author} {\bibfnamefont {J.}~\bibnamefont {Koralek}},
  \bibinfo {author} {\bibfnamefont {Y.}~\bibnamefont {Lu}}, \bibinfo {author}
  {\bibfnamefont {A.}~\bibnamefont {Vishwanath}}, \bibinfo {author}
  {\bibfnamefont {J.}~\bibnamefont {Orenstein}}, \bibinfo {author}
  {\bibfnamefont {D.}~\bibnamefont {Bonn}}, \bibinfo {author} {\bibfnamefont
  {W.}~\bibnamefont {Hardy}}, \ and\ \bibinfo {author} {\bibfnamefont
  {R.}~\bibnamefont {Liang}},\ }\bibfield  {title} {\enquote {\bibinfo {title}
  {New collective mode in \ce{YBa2Cu3O_{6+x}} observed by time-domain
  reflectometry},}\ }\href@noop {} {\bibfield  {journal} {\bibinfo  {journal}
  {Physical Review B}\ }\textbf {\bibinfo {volume} {88}},\ \bibinfo {pages}
  {060508} (\bibinfo {year} {2013})}\BibitemShut {NoStop}%
\bibitem [{\citenamefont {de~la Torre}\ \emph {et~al.}(2014)\citenamefont
  {de~la Torre}, \citenamefont {Hunter}, \citenamefont {Subedi}, \citenamefont
  {Walker}, \citenamefont {Tamai}, \citenamefont {Kim}, \citenamefont {Hoesch},
  \citenamefont {Perry}, \citenamefont {Georges},\ and\ \citenamefont
  {Baumberger}}]{de2014coherent}%
  \BibitemOpen
  \bibfield  {author} {\bibinfo {author} {\bibfnamefont {A.}~\bibnamefont
  {de~la Torre}}, \bibinfo {author} {\bibfnamefont {E.}~\bibnamefont {Hunter}},
  \bibinfo {author} {\bibfnamefont {A.}~\bibnamefont {Subedi}}, \bibinfo
  {author} {\bibfnamefont {S.~M.}\ \bibnamefont {Walker}}, \bibinfo {author}
  {\bibfnamefont {A.}~\bibnamefont {Tamai}}, \bibinfo {author} {\bibfnamefont
  {T.}~\bibnamefont {Kim}}, \bibinfo {author} {\bibfnamefont {M.}~\bibnamefont
  {Hoesch}}, \bibinfo {author} {\bibfnamefont {R.~S.}\ \bibnamefont {Perry}},
  \bibinfo {author} {\bibfnamefont {A.}~\bibnamefont {Georges}}, \ and\
  \bibinfo {author} {\bibfnamefont {F.}~\bibnamefont {Baumberger}},\ }\bibfield
   {title} {\enquote {\bibinfo {title} {Coherent quasiparticles with a small
  {Fermi} surface in lightly doped \ce{Sr3Ir2O7}},}\ }\href@noop {} {\bibfield
  {journal} {\bibinfo  {journal} {Physical Review Letters}\ }\textbf {\bibinfo
  {volume} {113}},\ \bibinfo {pages} {256402} (\bibinfo {year}
  {2014})}\BibitemShut {NoStop}%
\end{thebibliography}%


%

\begin{figure}
\includegraphics[scale=0.9]{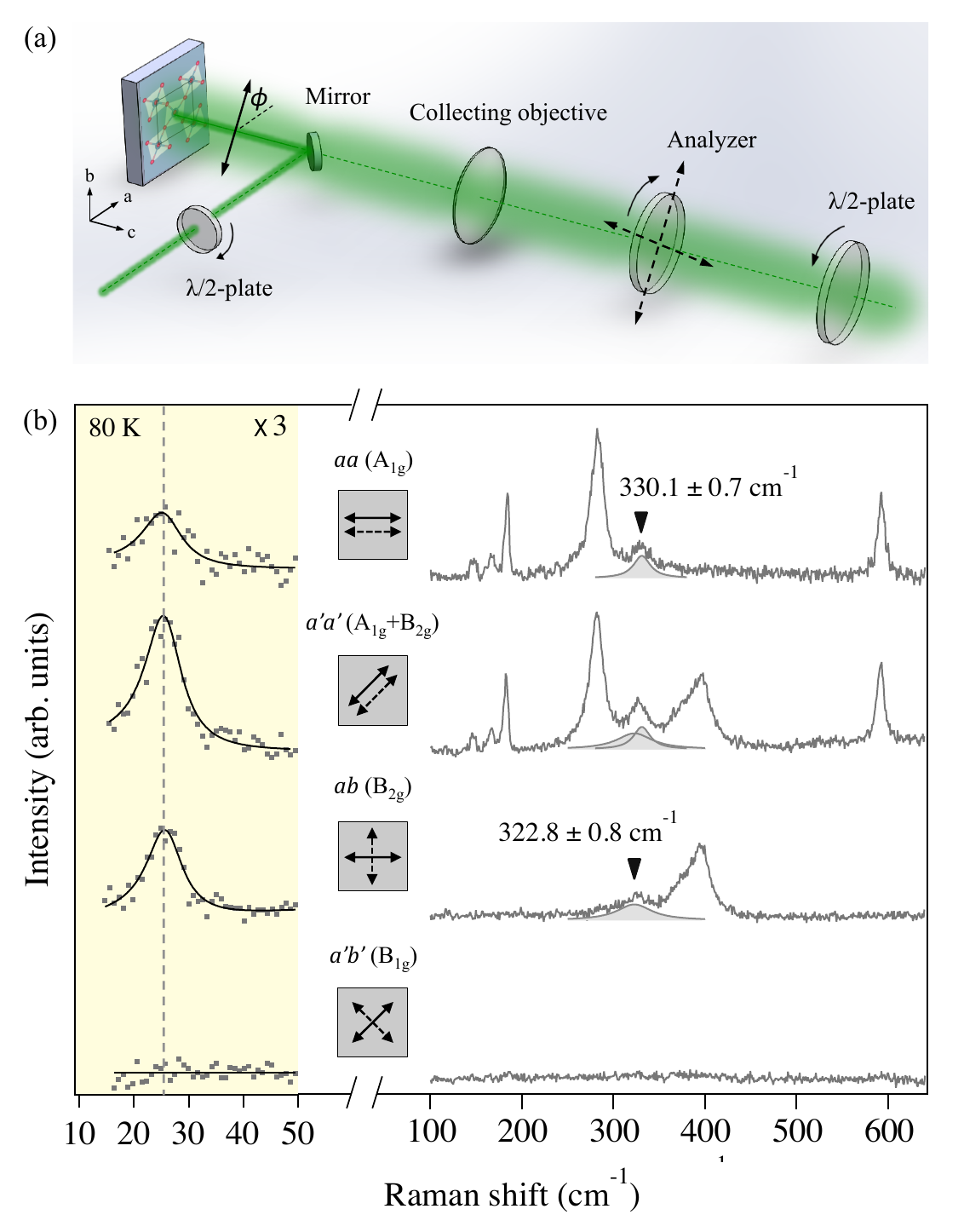}
\caption{\label{fig:fig.2} (a) Schematic of the rotation anisotropic polarized Raman spectroscopy experiment. The incident and scattered light propagates along the crystalline \textit{c} axis, and their polarizations are varied within the \textit{ab} plane. The polarization of the incident light (solid double arrows) is tuned by a half-wave plate, and the polarization of the scattered light (dashed double arrows) is selected by an analyzer to be either parallel or perpendicular to that of the incident light. The selected scattered polarization is then rotated back to vertical by a half-wave plate before sent into the spectrometer. All these optical components are strain-free to minimize the light depolarization. (b) Raman spectra of \ce{(Sr_{1-x}La_{x})_{3}Ir2O7} (x = 0.075) acquired in \textit{aa}, $a^{\prime}a^{\prime}$, \textit{ab}, and $a^{\prime}b^{\prime}$ channels at 80 K. The insets indicate the selection rule channels and the corresponding selected symmetry modes under $D_{4h}$ point group. The spectra in the low-frequency region is multiplied by a factor of 3 and highlighted in yellow shadow. Black curves are fits to the one oscillator function. The uncertainty represents +/- one standard deviation of the fitted frequency. The shaded profiles in \textit{aa} and \textit{ab} channels are Lorentzian fits to the modes marked by the black triangles, and the superposition of these two profiles accounts for the mode near \SI{330} {cm^{-1}} in the $a^{\prime}a^{\prime}$ channel.}
\end{figure}

\begin{figure}
\includegraphics[scale=0.5]{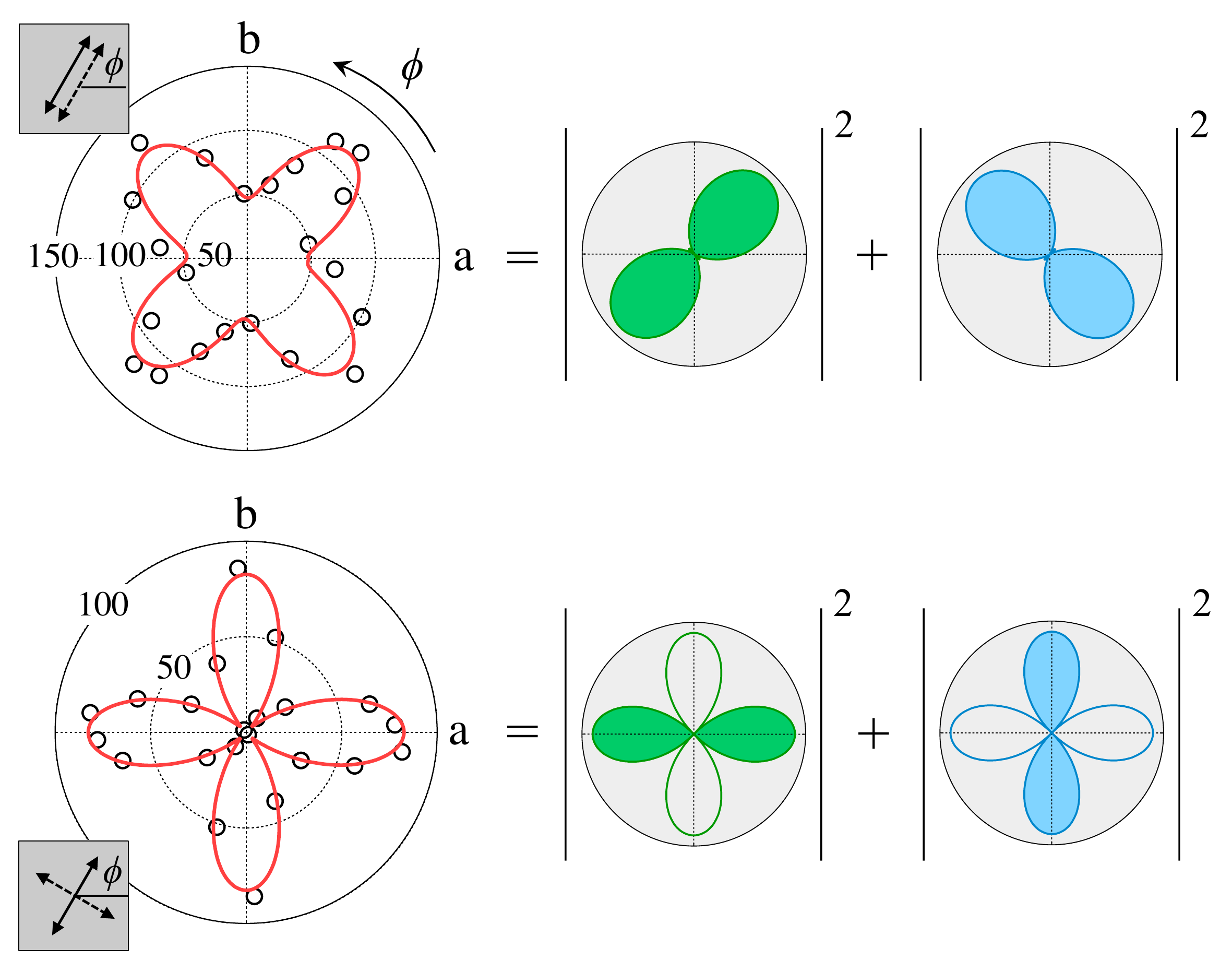}
\caption{\label{fig:fig.2} Polar plots of the fitted peak intensities of the charge order mode as a function of the rotation angle $\phi$ in \ce{(Sr_{1-x}La_{x})_{3}Ir2O7} (x = 0.075) at 80 K in the parallel (top) and perpendicular (bottom) configurations. The RA patterns are fitted to the sum of the Raman intensities from two domains with \SI{90}{\degree} rotated charge order wave vectors. The solid (empty) petals denote a positive (negative) phase for the scattered electric field. The error bars, $+/-$ one standard deviation of the fitted peak intensities, are smaller than the symbol sizes and therefore are not visible in the polar plots.}
\end{figure}

\begin{figure}
\includegraphics[scale=0.85]{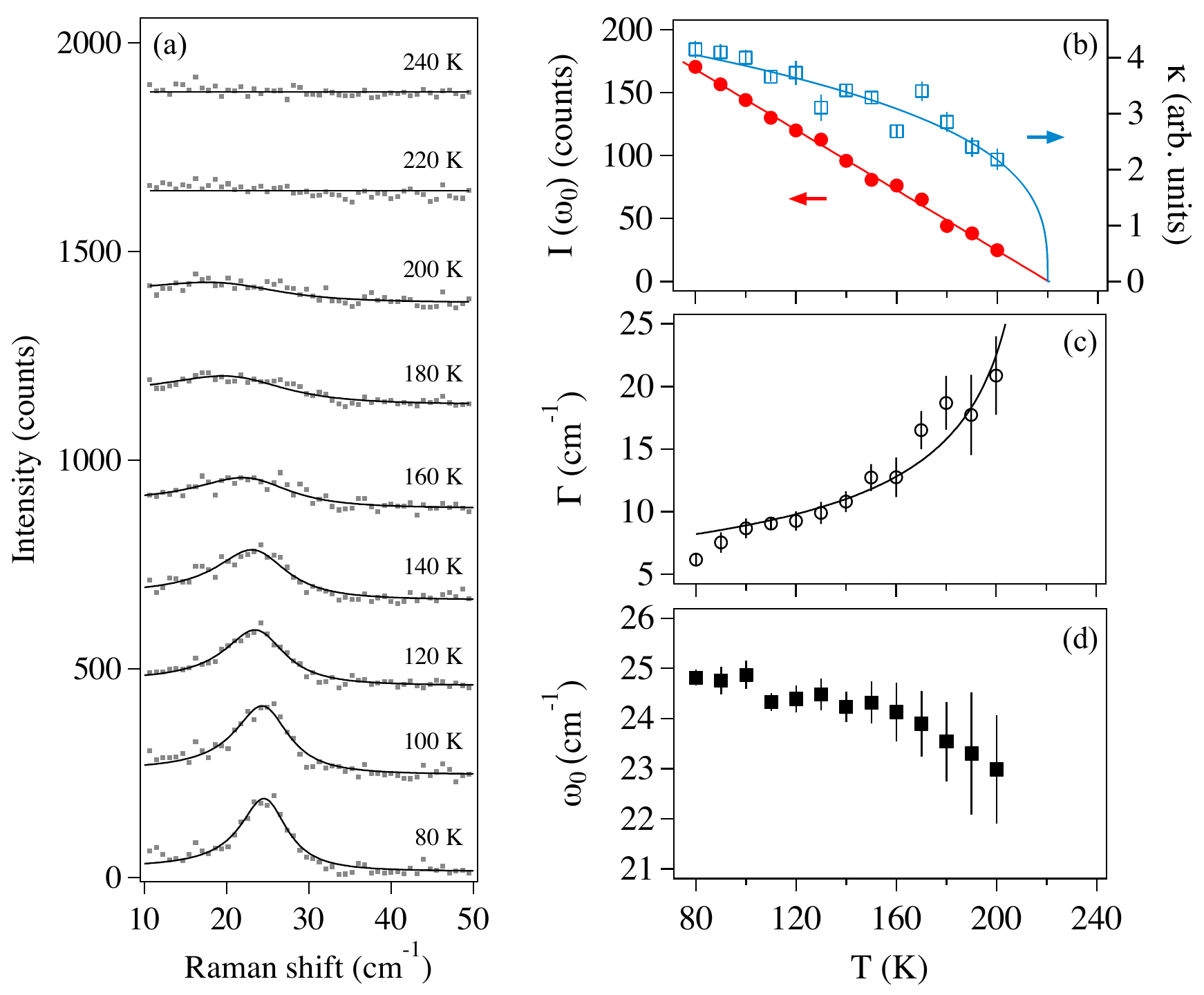}
\caption{\label{fig:fig.3} (a) Temperature dependence of the charge order mode in \ce{(Sr_{1-x}La_{x})_{3}Ir2O7} (x = 0.032). The spectra are vertically offset for clarity. The gray dots are raw data and solid curves are fits to the one oscillator function. (b)-(d) Temperature dependence of the peak intensity ($I (\omega_0)$, red solid circles), the coupling constant ($\kappa$, blue empty squares), the damping rate ($\Gamma$, black empty circles) and the frequency ($\omega_0$, black solid squares) of the charge order mode extracted from fits to the one oscillator function. Red, blue and black solid curves are fits to $I (\omega_0)=I_0(T_\mathrm{CO}-T)$ for the peak intensity, $\kappa=\kappa_0{(T_\mathrm{CO}-T)}^{\varsigma}$ for the coupling constant, and $\Gamma = \Gamma_0(T_\mathrm{CO} - T)^{-1/2}$ for the damping rate, respectively. The error bars are set to be +/- one standard deviation of the fitted parameters.}
\end{figure}

\begin{figure}
\includegraphics[scale=0.65]{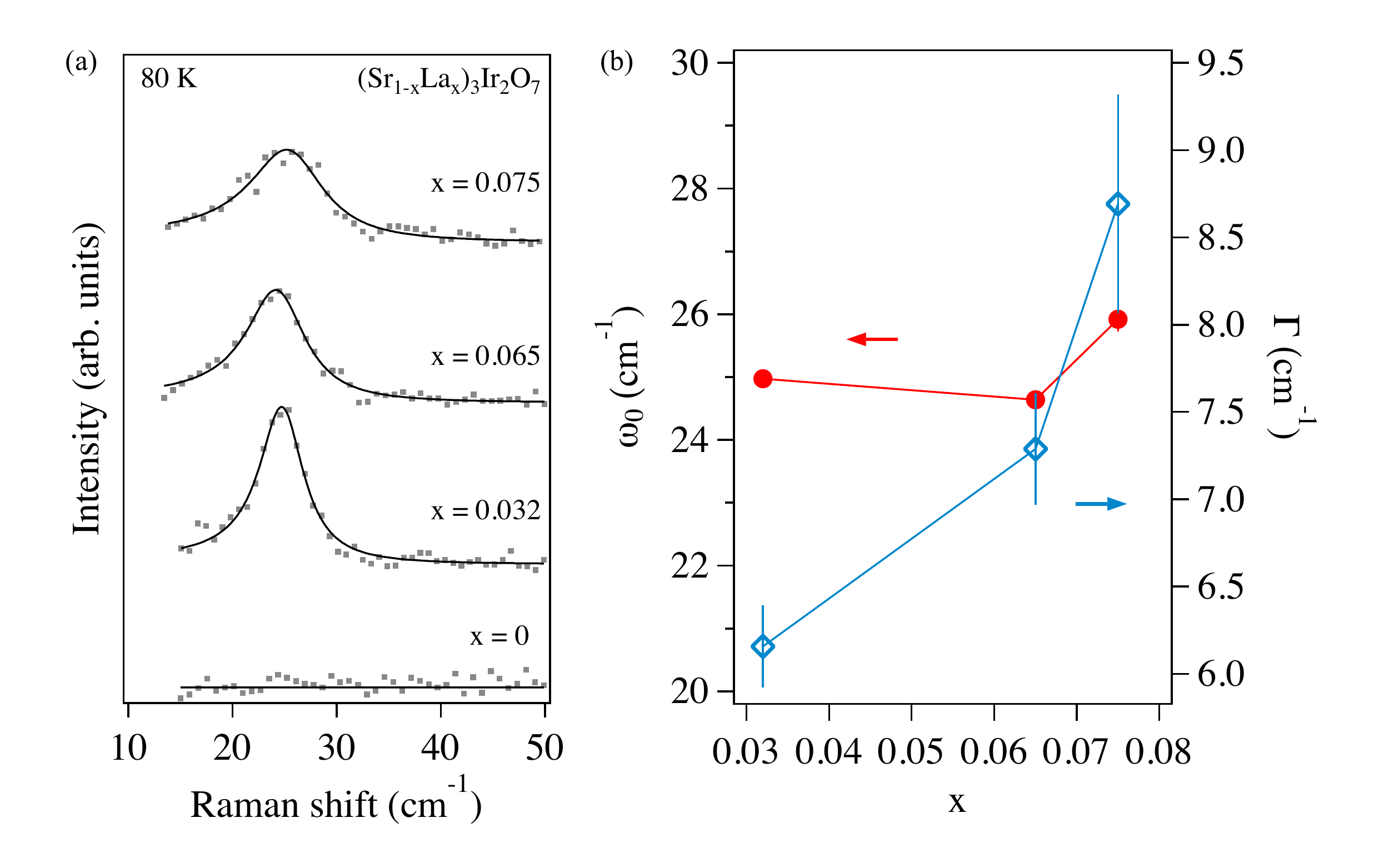}
\caption{\label{fig:fig.4} (a) Raman spectra of the charge order mode at 80 K in \ce{(Sr_{1-x}La_{x})_{3}Ir2O7} with x = 0, 0.032, 0.065, and 0.075, respectively. The spectra are acquired without an analyzer and are vertically offset for clarity. (b) Doping dependence of the charge order frequency ($\omega_0$, red solid circles) and the damping rate ($\Gamma$, blue empty diamonds) at 80 K. The error bars are set to be +/- one standard deviation of the fitted parameters.}
\end{figure}

\clearpage
\newpage

\begin{center}
\textbf{SUPPLEMENTARY MATERIAL\\}
\end{center}

\vspace{10pt}

\noindent\textbf{1. Survey of the Raman tensors of the $D_{4h}$ subgroups}\\
Table. I summerizes the Raman tensors of the orthorhombic subgroups ($D_{2h}$, $C_{2v}$, and $D_{2}$) of $D_{4h}$ point group. Given a Raman tensor (R) that has finite tensor elements in the \textit{aa}, \textit{ab} and $a^{\prime}a^{\prime}$ channels, but is fully suppressed in $a^{\prime}b^{\prime}$ channel, with the relationship of Raman intensity in the backscattering geometry $I = |\langle E_{I} | R | E_{S}\rangle|^2$, we can retrieve the Raman tensor form to be of $R=\begin{pmatrix} a & b & \\ b & a & \\  &  & c \end{pmatrix}$. By choosing a new coordinate system that is $\SI{45}{\degree}$ rotated about the \textit{c} axis, the Raman tensor is transformed into $R=\begin{pmatrix} a-b &  & \\  & a+b & \\  &  & c \end{pmatrix}$, matching the $A$ symmetry mode $(A_g, A_1, A)$ of the orthorhombic point groups.

\begin{table}[!h]
\caption{\label{table:table 1} Raman tensors of the orthorhombic subgroups.}
\renewcommand
\renewcommand
\centering
\begin{tabular} { c c c c c } 
\\
\hline
& $\begin{pmatrix} a &  & \\  & b & \\  &  & c \end{pmatrix}$ & $\begin{pmatrix}  & d & \\ d &  & \\  &  &  \end{pmatrix}$ & $\begin{pmatrix}  &  & e\\  &  & \\ e &  &  \end{pmatrix}$  & $\begin{pmatrix}  &  & \\  &  & f\\  & f &  \end{pmatrix}$ \\  [4ex]
 \hline
 $D_2$ & A & $B_1(z)$ & $B_2(y)$ & $B_3(x)$ \\
 \hline
 $C_{2v}$ & $A_1(z)$ & $A_2$ & $B_1(x)$ & $B_2(y)$\\ 
 \hline
 $D_{2h}$ & $A_g$ & $B_{1g}$ & $B_{2g}$ & $B_{3g}$\\ 
 \hline
\end{tabular}
\end{table} 

\vspace{10pt}

\noindent\textbf{2. Selection rules of Raman modes at room temperature}\\
The selection rules of Raman modes in \ce{(Sr_{1-x}La_{x})_{3}Ir2O7} (x = 0.075) at \SI{290}{K} in \textit{aa}, $a^{\prime}a^{\prime}$, \textit{ab}, and $a^{\prime}b^{\prime}$ channels are summerized in Fig. S1. The charge order mode near \SI{25}{cm^{-1}} is absent in all four channels. All the modes can be assigned to $A_{1g}$ or $B_{2g}$ symmetry mode of $D_{4h}$ point group, consistent with the reported results in Ref. [31] of the main text.\\

\newpage
\setcounter{figure}{0}  
\begin{figure}[!h]
\includegraphics[scale=0.75]{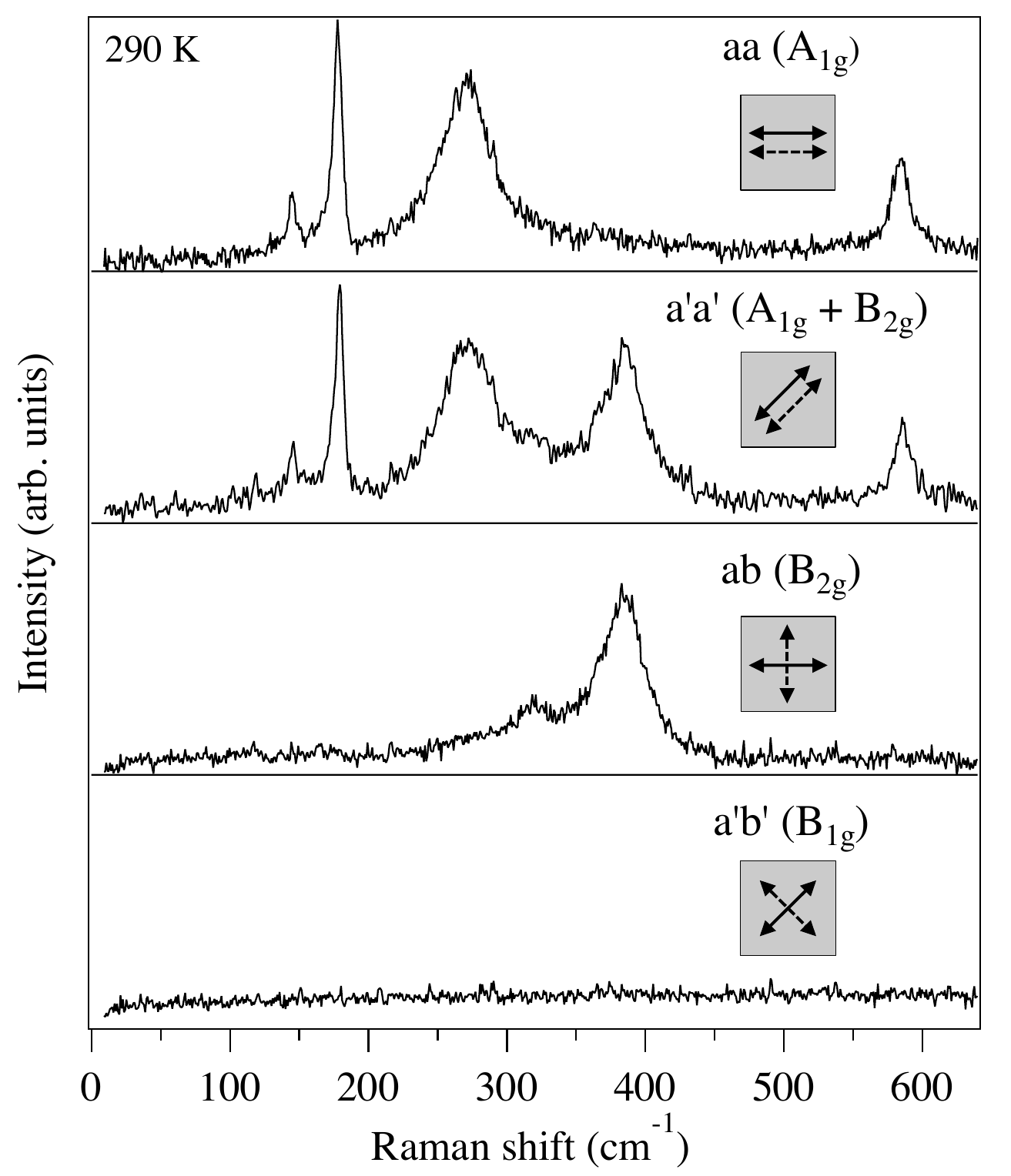}
\renewcommand{\thefigure}{S\arabic{figure}}
\caption{Selection rules of Raman modes in \ce{(Sr_{1-x}La_{x})_{3}Ir2O7} (x = 0.075) at \SI{290}{K}. The insets show the selection rules channels and corresponding selected symmtry modes in $D_{4h}$ point group.}
\end{figure}

\end{document}